\documentclass{pasj01}
\draft 
\Received{$\langle$reception date$\rangle$}
\Accepted{$\langle$acception date$\rangle$}
\Published{$\langle$publication date$\rangle$}

\Published{$\langle$publication date$\rangle$}

\usepackage{color}
\usepackage{ifthen}
\usepackage{natbib}
\usepackage{ulem}
\usepackage{multirow}
\usepackage{xcolor}
\usepackage{setspace}

\usepackage{arydshln}
\newcommand{\nside}{\ensuremath{N_{\rm side}}}
\newcommand{\kt}{\ensuremath{kT}}
\newcommand{\zo}{Z_{\odot}}
\newcommand{\zfe}{Z_{Fe}}
\newcommand{\fgal}{f_{\rm Gal}}
\newcommand{\nh}{\ensuremath{N_\mathrm{H}}}
\newcommand{\nhgal}{\ensuremath{N_\mathrm{H,GAL}}}
\newcommand{\Lend}{0.7}

\newcommand{\Txe}{\ensuremath{T_\mathrm{d}\ensuremath{(x, E)}}}
\newcommand{\Tcxe}{\ensuremath{T_\mathrm{d}^{\prime}\ensuremath{(x, E)}}}

\newcommand{\Cxe}{\ensuremath{C_\mathrm{d}\ensuremath{(x, E)}}}
\newcommand{\Hxe}{\ensuremath{H_\mathrm{d}\ensuremath{(x, E)}}}
\newcommand{\Nxe}{\ensuremath{N_\mathrm{d}\ensuremath{(x, E)}}}

\newcommand{\Nie}{\ensuremath{N_\mathrm{s}\ensuremath{(\vec{i}, E)}}}
\newcommand{\Nsie}{\ensuremath{N_\mathrm{s}^{\prime}\ensuremath{(\vec{i}, E)}}}

\newcommand{\Eix}{\ensuremath{E_\mathrm{x}\ensuremath{(\vec{i}, x)}}}

\begin{document}

%\SetRunningHead{Nakahira et al.}{diffuse structure observed with MAXI/SSC}

\title{MAXI/SSC All-sky maps from 0.7 keV to 4 keV}

\author{
	Satoshi \textsc{Nakahira}\altaffilmark{1,2},
	Hiroshi \textsc{Tsunemi}\altaffilmark{3},
	Hiroshi \textsc{Tomida}\altaffilmark{1},
  	Shinya \textsc{Nakashima}\altaffilmark{2},
   	Ryuho \textsc{Kataoka}\altaffilmark{4,5}
 and
   	Kazuo \textsc{Makishima}\altaffilmark{6},
	}

\altaffiltext{1}{Institute of Space and Astronautical Science (ISAS), Japan Aerospace Exploration Agency (JAXA), 3-1-1 Yoshinodai, Chuo, Sagamihara, Kanagawa, 252-5210, Japan}
\altaffiltext{2}{High Energy Astrophysics Laboratory, RIKEN, 2-1, Hirosawa, Wako, Saitama 351-0198, Japan}
\altaffiltext{3}{Department of Earth and Space Science, Graduate School of Science, Osaka University, 1-1
Machikaneyama, Toyonaka, Osaka 560-0043, Japan}
\altaffiltext{4}{National Institute of Polar Research, 10-3 Midori-cho, Tachikawa, Tokyo 190-8518, Japan.}
\altaffiltext{5}{SOKENDAI, 10-3 Midori-cho, Tachikawa, Tokyo 190-8518, Japan.}
\altaffiltext{6}{Kavli Institute for the Physics and Mathematics of the Universe (WPI), The University of Tokyo, 5-1-5 Kashiwanoha, Kashiwa, Chiba 277-8583, Japan}

% * <s.nakahira@gmail.com> 2017-11-10T05:34:29.566Z:
%
% ^.
%\email{***@***.***.***}


\maketitle

\begin{abstract}
By accumulating data from the Solid-state Slit Camera (SSC)
on board the MAXI mission for 2 years from 2009 to 2011,
diffuse X-ray background maps were obtained
in energies  of \Lend--1.0, 1.0--2.0, and 2.0--4.0 keV.
They are the first ones 
that were derived with a solid-state instrument,
and to be compared with the previous  ROSAT all sky survey result.
While the SSC map in the highest energy band is 
dominated by point sources and the Galactic Diffuse X-ray  emission,
that in \Lend--1.0 keV reveals an extended X-ray structure,
of which the brightness distribution is very similar 
to that observed with ROSAT about 20 years before.
Like in the ROSAT result,
the emission is dominated by a bright arc-like structure,
which appears to be a part of a circle
of $\sim 50^\circ$ radius
centered at about $(l, b) \sim (340^\circ, 15^\circ)$.
In addition, the SSC map suggests
a fainter and larger ellipse,
which is elongated in the north-south direction and
roughly centered at the Galactic center. 
The spectrum of these structures is explained 
as thin thermal emission from a plasma,
with a temperature of $\sim 0.31$ keV
and an abundance of  $\sim0.3$ Solar.
Based on SSC observation conditions
including the low Solar activity,
the Solar Wind Charge Exchange signals are estimated 
to be negligible in the present SSC maps,
as well as in the $>0.56$ keV ROSAT map.
A brief discussion is given to the obtained results.
\end{abstract}

\section{Introduction}

The ROSAT All Sky Survey (RASS) presented an entire sky map 
of the very soft X-ray background in the energy range of 0.1--2.5\,keV.
The map reveals complicated spatial structures 
which depend on the energy band. 
However, \citet{doi:10.1029/96GL03780} found 
that the very soft X-ray background was in fact heavily contaminated, particularly in the energy below 1\,keV, 
by many emission lines that arise through Solar Wind Charge Exchange (SWCX) 
in the Earth's vicinity \citep[for a review of the SWCX, see][]{2019A&ARv..27....1K}.
The SWCX signals are an astrophysical nuisance, 
because it is time-variable and depends on the direction, 
in such ways as are not yet well understood.  
This difficulty partly arose from the limited energy resolution of the RASS, 
which employed proportional counters.
Actually, observations of some sky regions with 
solid-state detectors were found to be much more
effective in resolving these low-energy emission lines
and separating the SWCX component 
\citep{2007PASJ...59S.133F,2007ApJ...661..304H}.
Yet another issue with the RASS was
that it was performed in 1990-1991,
when the Solar activity was very high.

In this way, a soft X-ray all-sky survey with a solid state instrument,
to be performed during a lower Solar activity, has been awaited.
After the RASS, 
several X-ray satellites \citep[ASCA, Chandra, XMM-Newton, and Suzaku][]{}
actually observed the soft X-ray sky in energy ranges below 2\,keV,
employing X-ray CCDs with a much better energy resolution than that of RASS.
However, they were basically pointing instruments
coupled to focusing X-ray telescopes
with a relatively narrow field of view (FOV).
As a result,
they were unable to uniformly map the whole sky.  

MAXI \citep[Monitor of All-sky Image;][]{2009PASJ...61..999M} 
is an all sky X-ray imager, 
launched in 2009 and mounted onboard the international space station (ISS).  
Of the two scientific instruments of MAXI, 
the SSC \citep[Solid-state Slit Camera;][] 
{2010PASJ...62.1371T, 2011PASJ...63..397T} is a pair of  X-ray CCD cameras
which are sensitive in an energy range of  \Lend--7 keV. 
Because the SSC uniformly covers the entire sky, 
it is expected to provide the first opportunity 
to study the all-sky soft X-ray background
with a solid-state energy resolution,
and to follow up the RASS result.
Another obvious advantage of using the SSC is 
that the Solar activity was very low in the early phase of the MAXI mission,
so the SWCX signals must have been considerably lower than in the RASS period.
For reference, the other MAXI instrument,
the GSC \citep[Gas-Slit Camera;][]{2011PASJ...63S.623M, 2011PASJ...63S.635S},
is less suited for this purpose,
because it employs an array of large-area gas proportional counters,
and it works in a higher energy range above 2 keV.

In the 20 years since the RASS, a new scientific importance
has been added to the study of soft X-ray sky.
Namely, a giant Galactic gamma-ray structure was discovered 
by the Fermi Gamma-Ray Space Telescope, 
in the energy range of  1-10\,GeV \citep{2010ApJ...717..825D}.  
The feature emerges from the Galactic Center (hereafter GC), 
and extends symmetrically up to 50$^\circ$ north and south from the Galactic Plane. 
The dumbbell-like overall shape, called the Fermi Bubble, 
is suggestive of an explosion from the GC.  
Furthermore, several studies suggest \citep[e.g.][]{2013ApJ...779...57K} 
that the RASS X-ray feature is related, at least partially, to the Fermi Bubble.  
\citet{2013ApJ...779...57K} performed several pointings
onto the Fermi Bubble edge with the Suzaku XIS \citep{2007PASJ...59S..23K} 
which has a relatively small FOV of 18$^\prime$ square, 
and detected an intensity or temperature jump at the edge location.  
Thus, the study of extended soft X-ray structures will potentially
provide us with information on large-scale energetic phenomena 
in the Milky Way and the GC.

In the present paper, we describe the MAXI/SSC data reduction,
and modeling of its non X-ray background.  
Then, the obtained soft X-ray maps are presented 
in three energy bands,\Lend--1.0, 1.0--2.0, and 2.0--4.0 keV.
They are compare with those with ROSAT,
across the time interval of 20 years.  
After conducting spectral analyses of
brighter parts of the observed soft X-ray emission,
we compare the present results  with those from ROSAT.

\section{Instruments}\label{sec:inst}

The SSC consists of two units of identical CCD cameras, SSC-H and SSC-Z. 
One SSC unit has 16 CCDs in 2$\times$8 arrays, 
and each CCD is a 25\,mm square chip (24\,$\mu$m square pixel) 
with no storage area.  
In order to avoid optical light, 
the CCD surface is coated with aluminum of 0.2\,$\mu$m thick,
of which the transmission efficiency is 
63\% at 1\,keV for normal incidence.  
They are cooled below $-60^\circ$C by the combination 
of a radiator and a Peltier device,
and are running in a pixel-sum mode to be a one-dimensional imager.  
The FOV of each camera is restricted to 90$^{\circ}\times$3$^\circ$,
by a collimator made of 24 parallel Cr-coated bronze plates
and a slit made of two tungsten bars with sharp edges.  
SSC-H and SSC-Z have FOVs
which are pointing to the horizontal and the zenith directions, respectively,
and both scan over wide sky (30--40\% of the entire sky)
as the ISS rotates around the Earth with a period of 92 min.  
Meantime, each location of the sky is viewed by either camera
for a typical on-source time of 30--40\,s. 
The point spread function (PSF) of the SSC is 
approximated by a circle of $1^\circ$.5 radius.

Since the MAXI operation started in 2009 August, 
the CCDs have gradually been degrading 
due to radiation damages by high energy particles.  
The radiation dose is particularly high 
when the ISS gets through the South Atlantic Anomaly (SAA), 
and through high latitude regions 
since the ISS orbit has an inclination of about 51$^\circ$.6.  
From the data analysis point of view, 
the radiation damages on the SSC appear mainly 
in two properties \citep{2017symm.conf....9T}; 
one is degradation of the energy resolution 
which can be continuously monitored referring to Cu-K lines 
from the bronze plates of the collimator, 
and the other is an increase of background events 
due to thermal noise at low energies.  
Furthermore, these effects differ among the CCD chips, 
depending on their locations within the camera unit.  
The chips just below the slits have been subject to the heaviest damage, 
while those at the far-ends remained the least affected.  
At the beginning of the mission, 
all the CCDs showed similar performance, 
but their performance differences have gradually developed.

\section{Data analysis}

In this paper, 
we utilize the SSC data that were acquired 
for about 2 years from 2009 August 17 to 2011 August 27,
so that we can utilize the energy range down to \Lend\,keV 
without hampered by the increase of the low-energy background.  
These data were partly analyzed by \citet{2013PASJ...65...14K} 
in a preliminary way.  
In our analysis, we employ a new method to estimate 
the background prior to creating the all sky maps, 
and fully take into account the position dependence 
of the energy response matrix and of the background spectra.  
The quantum efficiency curves of the CCD were also calibrated 
using the Crab Nebula; 
the derived photon index of the \Lend--7\,keV Crab spectrum, 
previously $2.20\,\pm 0.05$ \citep{2010PASJ...62.1371T}, 
has been revised to $2.10\pm 0.05$, 
which agrees better with the canonical value of 
$2.125\,\pm 0.001$ \citep{2013PASJ...65...74K}.

\subsection{Data Reduction}

The live time of the SSC observations for the above time period, 
with standard data-screening criteria \citep{2010PASJ...62.1371T}, 
amounted to 2.61$\times$10$^7$ s, or 41\% of real time.  
Since these data are dominated by the backgrounds, 
good time intervals must be carefully selected.  
In the high latitude regions above $\pm$40$^\circ$, 
the SSC data occasionally showed flaring behavior, 
probably induced by geomagnetic particle activities.  
Hence, we examined the Radiation Belt Monitor data 
of the MAXI/GSC \citep{2011PASJ...63S.635S}, 
and removed the time intervals of increased particle counts.  
In the SSC, very bright infrared light leaves 
light-leakage effects \citep{2010PASJ...62.1371T}.  
Therefore, the data were discarded 
when the Sun elevation angle was higher than $-10^\circ$ from the Earth horizon,
or the Moon was within $10^\circ$ of the FOV.  
These screenings have reduced the live time to 
$1.51 \times 10^7$ s (23.6\% of the real time).  
In addition, data with sporadic background increases were excluded, 
by removing noisy pixels, columns, or areas, 
which are defined for each orbit.  
Through all these screenings, 
we have been left with $1.35\times10^7$ and $0.57\times10^7$ photons, 
in the \Lend--7\,keV and 7--9\,keV energy bands, respectively.  

\begin{figure}
 \begin{center}
      \FigureFile(80mm, 80mm){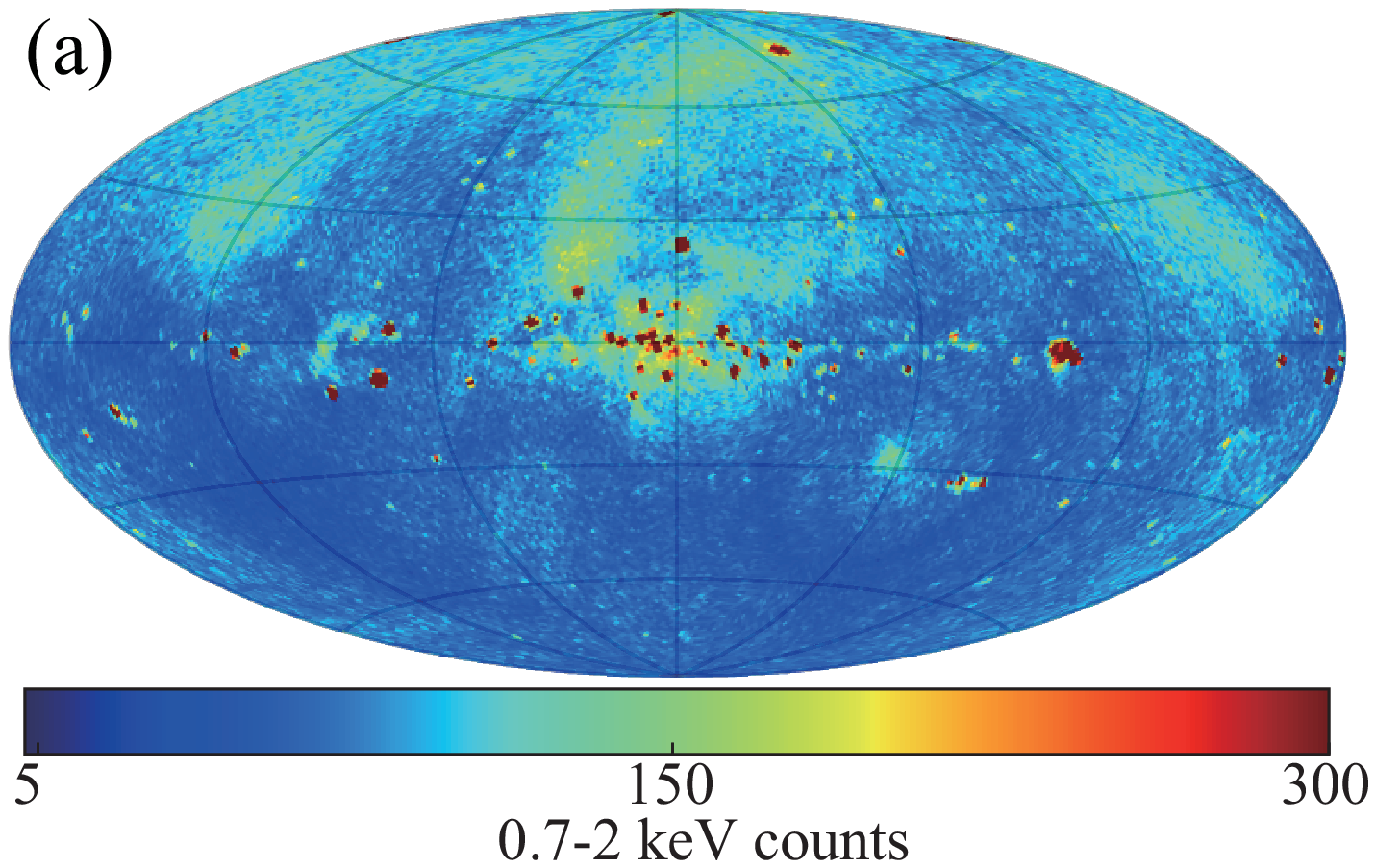}
      \FigureFile(80mm, 80mm){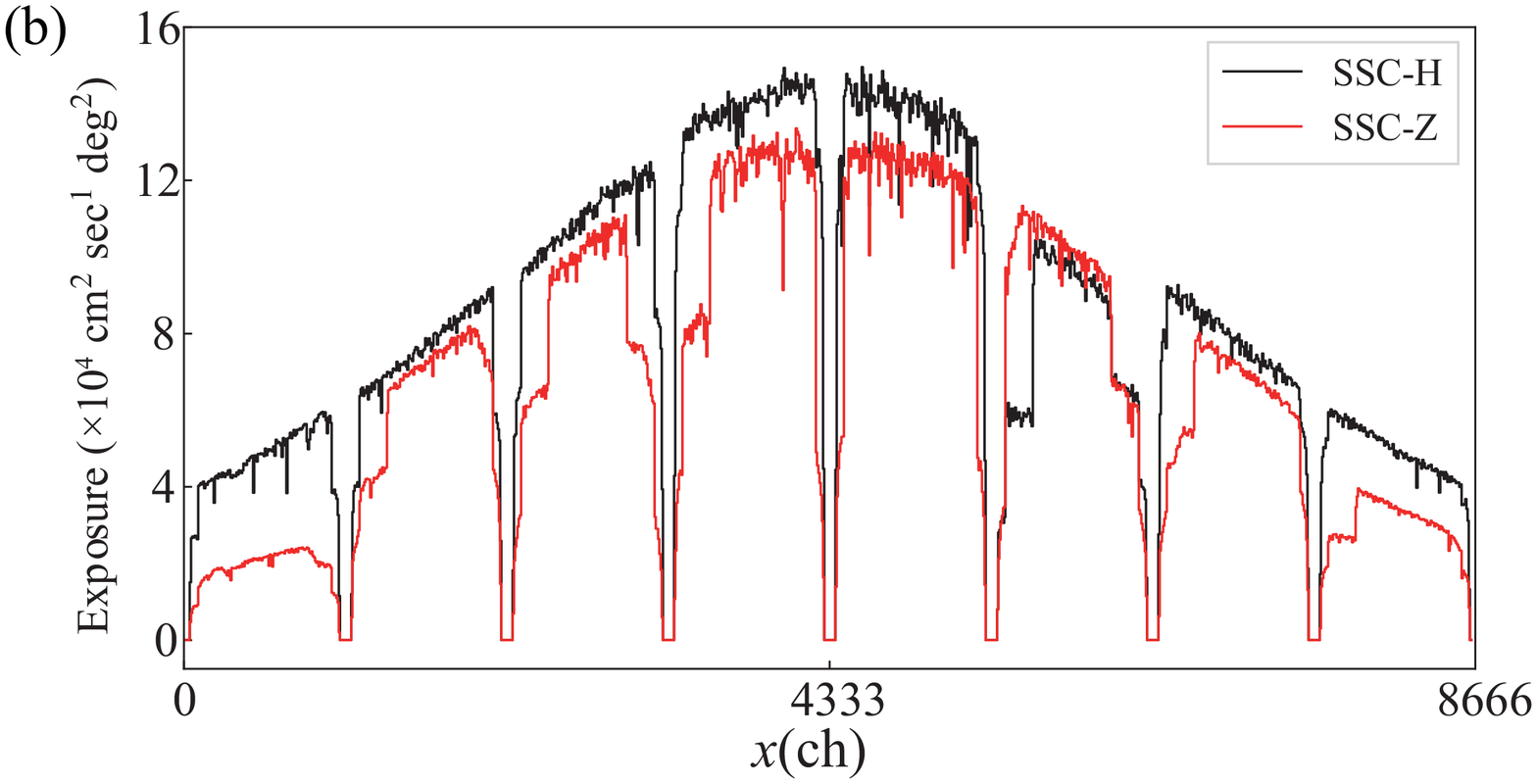}
      \FigureFile(80mm, 80mm){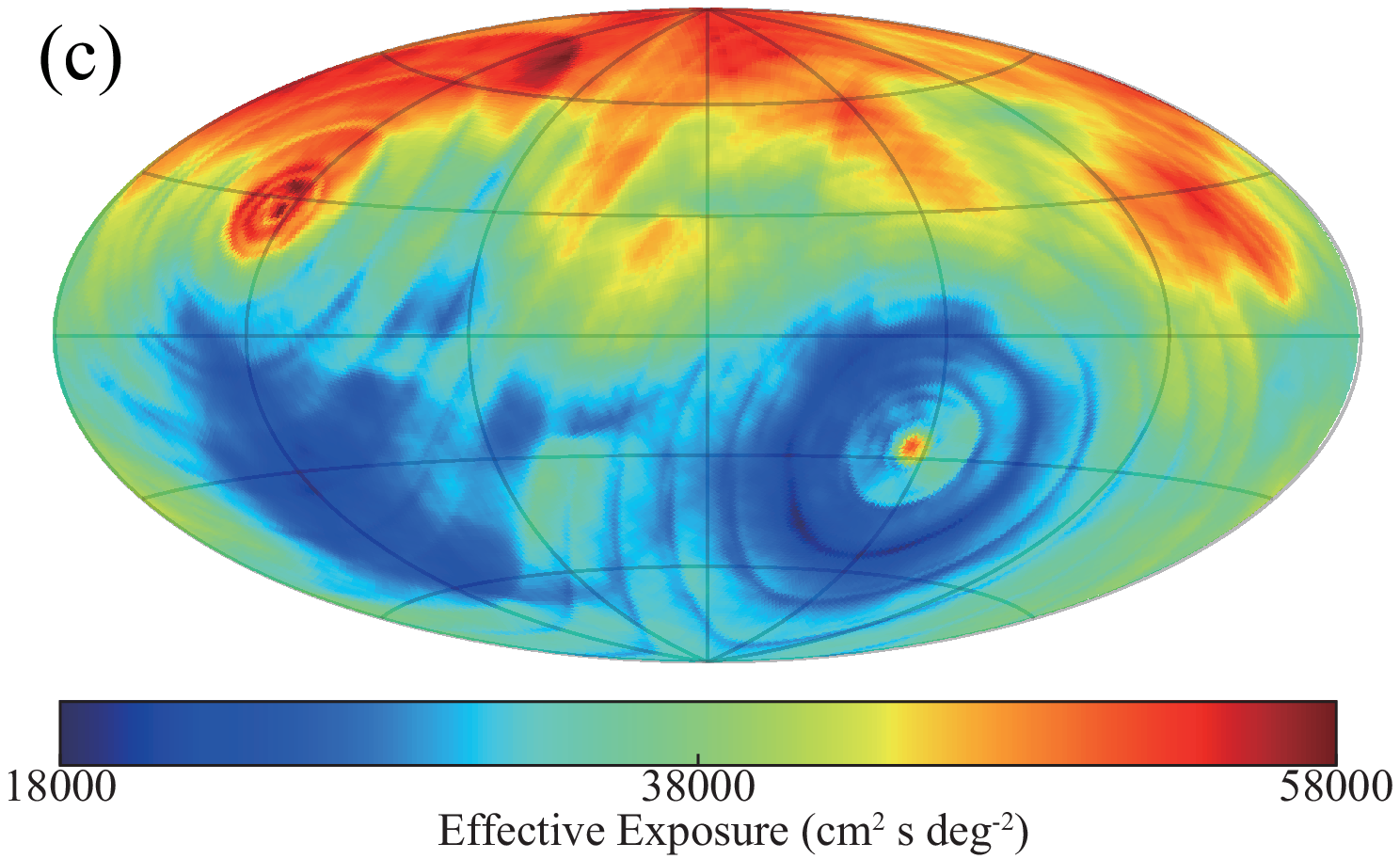}
 \end{center}
 \caption{
(a) \Lend-2 keV all-sky image obtained with the MAXI/SSC,  
presented in the Galactic coordinates.  
All the data screenings are applied 
while it is not corrected for the exposure.  
(b) The SSC exposure maps projected 
onto the detector coordinate $x$.  
Black and red indicate the data from SSC-H and SSC-Z, respectively.  
(c) The exposure map, with SSC-H and SSC-Z summed, 
projected onto the sky coordinate, $\vec{i}$.
} \label{expomaps}
\end{figure}

Figure\,\ref{expomaps}(a) shows the \Lend--2 keV all-sky image, 
using the events that survived the above screening criteria.  
It is presented in the Aitoff-Hammer projection and the Galactic coordinates,
which are hereafter employed throughout.  
Considering all the screening processes, 
and employing the {\tt HEALPix} software library 
\citep{2005ApJ...622..759G} 
where the number of pixels was substituted for the parameter $N_{\rm side}$,
the exposure maps $\Eix$ were also calculated, 
separately for SSC-H and SSC-Z.
They are expressed as a function 
of the pixel number of the sky (sky coordinate), $\vec{i}$, 
and the detector coordinate, $x$.
In other words, a particular sky region $\vec{i}$ was
integrated at a particular detector coordinate $x$
for a total exposure time of $\Eix$.
Since the CCDs of the SSC are used as one-dimensional imagers
(section \ref{sec:inst}), $x$ is a one-dimensional variable;
in the orthogonal direction, the CCD pixels are summed up. 
Figure\,\ref{expomaps} (b) and (c) present the projections 
of $\Eix$ onto the $x$ and $\vec{i}$ coordinates, respectively.  
The sharp drops periodically seen in panel (c) 
are gaps between the CCD chips.
In panel (c), the exposure per sky area, 
integrated over $x$,
scatters over 1.8-5.8 $\times$ 10$^4$ cm$^2$ s deg$^{-2}$, 
with a total exposure of 1.25$\times$10$^9$ cm$^2$ s deg$^2$ 
when the two camera units are coadded.

%%consequently they show narrow strips of missing data.

\subsection{SSC Backgrounds} 

The present paper focuses on the extended X-ray structure (EXS) 
in the diffuse soft X-ray emission,
which was revealed by the RASS in its 0.44--1.21 keV band (R4+5).
When studying this phenomenon, however,
we need to identify, estimate, and subtract 
all the other extended emission components 
that contribute to the SSC data.
Some of them are as extended as the EXS itself,
and some can be time variable.
Below, we collectively call all the other components as backgrounds,
and try to construct a method to estimate and remove them.
This makes a big difference from studies of point sources,
in which the background can be estimated from surrounding sky regions
\citep{2013ApJS..207...36H, 2016PASJ...68S..32T}.

\subsubsection{Background components}\label{bgcomponents}

The soft X-ray emission over the entire sky
is considered to consist mainly of five celestial components
with different spectral and/or spatial properties,
and two non-celestial ones.
Generally from larger to smaller physical scales, 
the former five are; the cosmic X-ray background (CXB), 
the Galactic halo emission (GH), 
the Galactic Diffuse X-ray emission (GDXE), 
the EXS which is the subject of the present study, 
and the emission from the local hot bubble (LHB).  
As to the latter, we have to consider
the non-X-ray background (NXB) and the SWCX.

The CXB is considered to be isotropic and temporally stable, 
with a power-law spectrum in the energy range below 10 keV.  
According to \citet{2002PASJ...54..327K} 
who used the ASCA GIS to study the CXB, 
its 2--10 keV spectrum is expressed by a power law 
of photon index $\Gamma = 1.412$, 
and its surface brightness fluctuates by 6.5\% (the 1-$\sigma$ value) 
within the GIS FOV (a circle of $0^\circ .35$ radius).  
Based on their results, we estimate the \Lend--2\,keV CXB surface brightness 
as $6.3\times 10^{-8}$ erg\,cm$^{-2}$\,s$^{-1}$\,sr$^{-1}$, 
and the brightness variations as 3\% and 0.3\% 
for the PSF of the SSC and a circle of $15^\circ$ radius, respectively.  
The GH component was studied by \citet{2018ApJ...862...34N} in detail 
using the Suzaku XIS data, mainly on the Galactic anti-center hemisphere.  
Its spectrum is expressed by a thin thermal plasma model 
with a temperature of $\sim 0.26$ keV.  
The emission measure (EM\footnote{EM = $\int n^2 dL$, 
where $n$ is the plasma density 
and $L$ is the plasma depth}) is found to be marginally correlated 
with cosec$|b|$ around its median of $0.31\times 10^{-2}$ cm$^{-6}$ pc,
but its global distribution is still unclear. In our analysis, 
we assume that the GH is uniform in temperature and intensity for simplicity.  

The GDXE comprises the Galactic-ridge X-ray emission \citep{1982ApJ...255..111W} 
and the Galactic-Center diffuse X-ray emission \citep{1989Natur.339..603K}, 
which are associated with the Galactic plane 
and a region within $\sim2^\circ$ of the GC, respectively.  
Although the GDXE, with its strong localization, 
would not provide a direct nuisance to our EXS study, 
we must carefully remove it when creating the model background 
which is to be subtracted from the raw SSC data 
(see $\S$\ref{subsubsec:bkdsubt}).  
The LHB emission is of thermal origin, 
characterized by a temperature of $kT\sim 10^6$\,K \citep{1997ApJ...485..125S}.
Since this temperature is too low to significantly contribute 
to the energy above \Lend\,keV, we neglect the LHB contribution.

In studying extended sources where the neighboring-sky subtraction 
\citep{2013ApJS..207...36H, 2016PASJ...68S..32T} cannot be used, 
the NXB is one of the largest nuisances. 
In other satellites in low-Earth orbits, 
like ASCA \citep{2002PASJ...54..327K} and Suzaku \citep{2008PASJ...60S..11T},
the NXB can be isolated and directly measured by the data accumulated 
when the instrument is pointing to the dark Earth.  
However, the ISS keeps the same attitude with respect to the Earth, 
and hence MAXI is always scanning the sky, 
except very infrequent events like the docking by the Soyuz spacecraft. 
Therefore, in our study of the EXS, 
we must  estimate and subtract the NXB
using the SSC data themselves
that are accumulated from sky directions.

Finally, we must also consider the SWCX phenomenon,
which was first noticed from comet observations 
by ROSAT \citep{doi:10.1029/96GL03780}.
This foreground emission will affect 
the soft X-ray sky with many emission lines,
particularly in energies below 1\,keV.
Compared to the RASS case,
this contamination is expected to be 
relatively insignificant in the present SSC data,
because the sunspot number was by an order of magnitude 
lower in 2009-2011 than in the ROSAT period,
and because we use the SSC data above \Lend\,keV 
where few SWCX lines are expected. 
For example, the SSC efficiency for the strongest SWCX line, 
O-VIII at 0.65\,keV, is only a few percent, 
which is less than 1/30 of that at 1\,keV.
Nevertheless, the SWCX is a phenomenon 
with sporadic time variability and complex directional anisotropy.
Therefore, its effects on the present data should be carefully evaluated.

\subsubsection{Modeling of the SSC backgrounds}
\label{subsubsec:bkdmodel}

\begin{figure}
 \begin{center}
  \includegraphics[width=80mm]{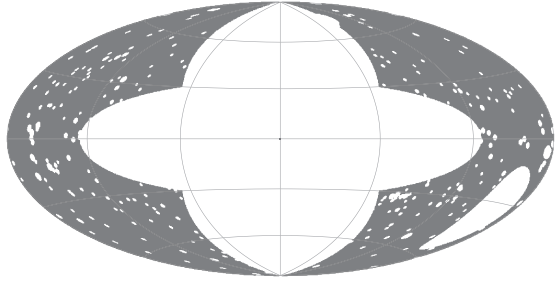}
 \end{center}
 \caption{The off-source region (the gray region),
 shown in the galactic coordinate. 
It was obtained by masking out the central regions
where the ROSAT EXS is significant,
and a number of remaining point sources (see text).
 Grid intervals are $60^\circ$ in longitude, 
 and $30^\circ$ in latitude. 
 }
\label{bgmasks}
\end{figure}

The gray region shown in figure\,\ref{bgmasks} is
what we call the off-source region,
where we attempt to determine the NXB model.  
It was obtained by excluding, from the entire sky,
two big elliptical regions shown in white;
one is along the Galactic plane where the GDXE is seen,
and the other wider one connecting the north and south poles
is the region where the EXS itself is significant. 
The latter approximates the region in 
the ROSAT R5 band map
\footnote{http://www.jb.man.ac.uk/research/cosmos/rosat/} \citep{1995ApJ...454..643S} 
where the soft X-ray brightness is above 
$75 \times10^{-6}$ counts cm$^{-2}$ s$^{-1}$.
The gray off-source region defined in this way covers 45\% of the sky.
Although the use of this off-source region is
primarily to subtract the NXB,
this subtraction process is expected to also remove,
at least approximately, the SWXC signals
that may have leaked into the present SSC data.
This is because the regions where the SWCX emission is strongest
(the cusps above the geomagnetic poles; see subsection \ref{subsec:discuss_swcx}),
will move over relatively wide sky regions
when projected onto the sky plane as seen from the ISS,
as the ISS makes its revolution,
and as the ISS orbital plane precesses 
around the Earth with a period of 70 days.
Further consideration of this issue continues in subsection \ref{subsec:discuss_swcx}.

\begin{figure}
 \begin{center}
    \FigureFile(80mm, 80mm){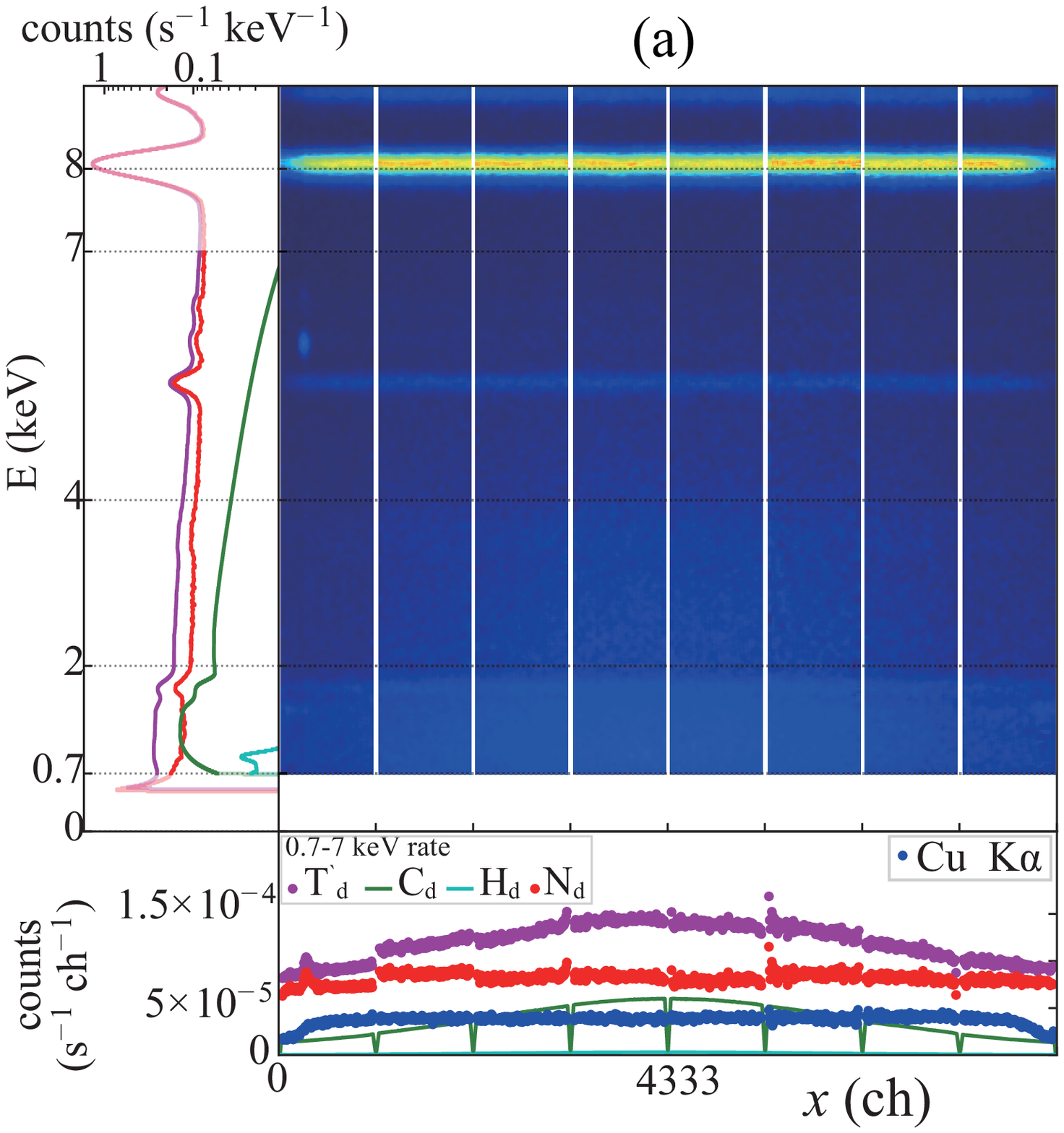}
    \FigureFile(80mm, 80mm){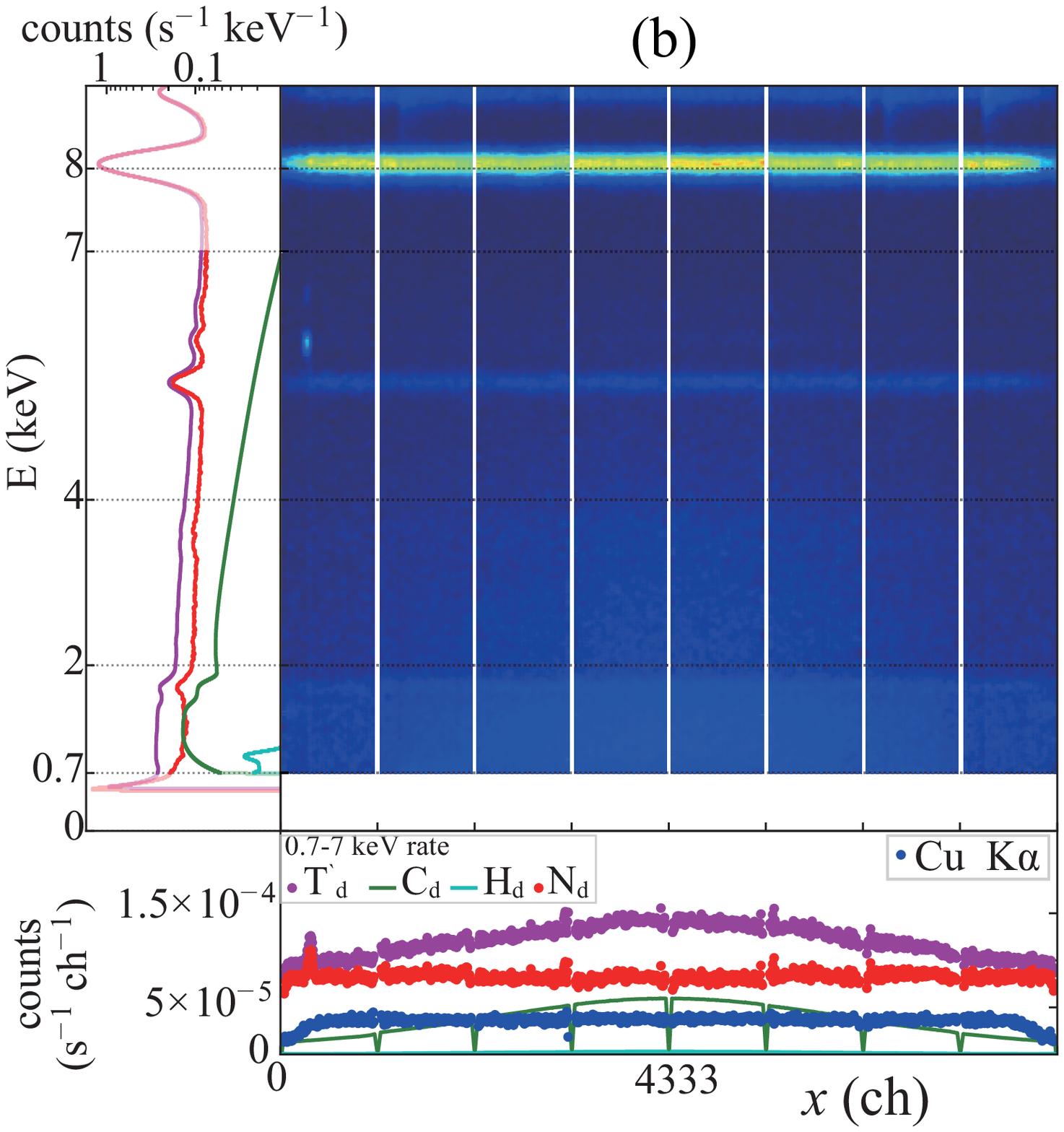}
 \end{center}
 \caption{
Background maps $T'(x,E)$ in the off-source region, 
accumulated over 2 years, for SSC-H (panel a) and SSC-Z (panel b). 
At the bottom and to the left of each map, 
the projections onto the $x$ axis (i.e.,the 1-dimensional profile) 
and the $E$ axis (i.e., X-ray spectrum) are shown, respectively.  
There, magenta shows the total background, 
green the calculated CXB contributions, 
cyan the GH, red the [total$-$(NXB+CXB)],
and blue (in the bottom panel) the Cu-K$\alpha$ line strength 
determined by a Gaussian fitting to the spectrum at each $x$.
}
\label{bgtemplates}
\end{figure}

By accumulating the SSC events over the above-defined off-source region, 
we obtained a total background map $\Txe$, 
as a function of $x$ 
and the photon energy $E$.  
Then, $\Txe$ was exposure-corrected to be $\Tcxe$, 
which is shown in figures\,\ref{bgtemplates}(a) for SSC-H, 
and figures\,\ref{bgtemplates}(b) for SSC-Z.
Each map is projected onto the $x$ axis 
to represent the positional photon distribution (bottom panel),
and onto the $E$ axis to provide the energy spectrum (left panel).  
In these maps, the CCD gaps are again seen as
vertical white stripes every 1024 pixels in $x$.
We also see two emission lines running horizontally on each map; 
a strong Cu-K line (8.04\,keV) and a weak Cr-K line (5.41\,keV), 
both generated via fluorescence on the SSC collimator plates, 
mainly by charged particles 
and partly by high energy X-rays and gamma-rays.  
A very weak $^{55}$Fe source (5.9\,keV) is also seen 
at the left end of this figure.  
As revealed by the $x$-dependence of the total events, 
shown in a purple line, 
the center CCDs which are closer to the SSC slit receive 
more photons (mainly the CXB)  than the side ones.

The background thus extracted from the off-source region 
can be regarded as consisting mainly of the CXB, the GH, and the NXB,
with a possible addition of the SWCX signals.
These components (possibly except SWCX) are 
considered to be relatively uniformly distributed over the sky.
Even if the brightness of any of them vary 
to some extent across the off-source region,
the variation would not introduce a significant 
$x$-dependence in figures\,\ref{bgtemplates}
(except the $x$-dependent instrumental responses),
because each $x$ is considered to have rather uniformly 
sampled various positions of the off-source region
through the 2-year observation.
Given these, we calculated the expected CXB and GH 
contributions to the SSC background, 
denoted as $\Cxe$ and $\Hxe$ respectively,
from the assumed spectral models and the energy response function.  
We employed the spectral model for the CXB as already described.  
The GH emission was represented by 
an {\tt APEC} plasma model in {\tt Xspec},
with the plasma temperature of $\kt=0.26\,$ keV, 
the metallicity of $Z= 1\zo$,
and an emission measure of $0.31\times10^{-2}$ cm$^{-6}$ pc 
\citep[median of][]{2018ApJ...862...34N}.  
In figure\,\ref{bgtemplates}, these components are shown 
by the green and cyan lines, respectively.

When $\Cxe$ and $\Hxe$ constructed in the off-source region
are removed from $\Tcxe$, the remaining component,
denoted as $\Nxe \equiv \Tcxe-\Cxe-\Hxe$
and shown in red in the bottom panels,
is quite constant as a function of $x$, 
except slight dependence on the CCD chips.  
Therefore, we conclude that $\Nxe$ is mainly the NXB, 
which is not coming from the celestial sphere,
and its chip dependence is probably due to 
the difference of the working temperature. 
We also find that the Cu-K line intensity (blue points) is very uniform, 
and almost chip independent, except at the very ends of the SSC; 
this is reasonable, because these line photons come from the collimator plates,
and their intensity is positively correlated with the NXB.
Hereafter, the term NXB is used to include these instrumental lines
as well as the SWCX photons if any.

%% We construct the background (CXB + GH + NXB) in the sky coordinate as below.  First of all, we integrate $\Cxe$, $\Hxe$ and $\Nxe$ as

\subsubsection{Subtraction of the modeled backgrounds}
\label{subsubsec:bkdsubt}

As described above, we used the off-source region
to estimated the NXB as a function of $x$.
We then attempt to transform it into the NXB spectrum,
$\Nie$, to be observed at a sky position $\vec{i}$,
not only in the off-source region but also in any sky direction.
This can be done by integrating $\Nxe$ over $x$ as
\begin{eqnarray*}
\Nie= \int \Eix\times\Nxe dx .
\end{eqnarray*}
This is a convolution procedure, 
in which $\Eix$ works as a transfer (or kernel) function.

\begin{figure}
 \begin{center}
    \FigureFile(80mm, 80mm){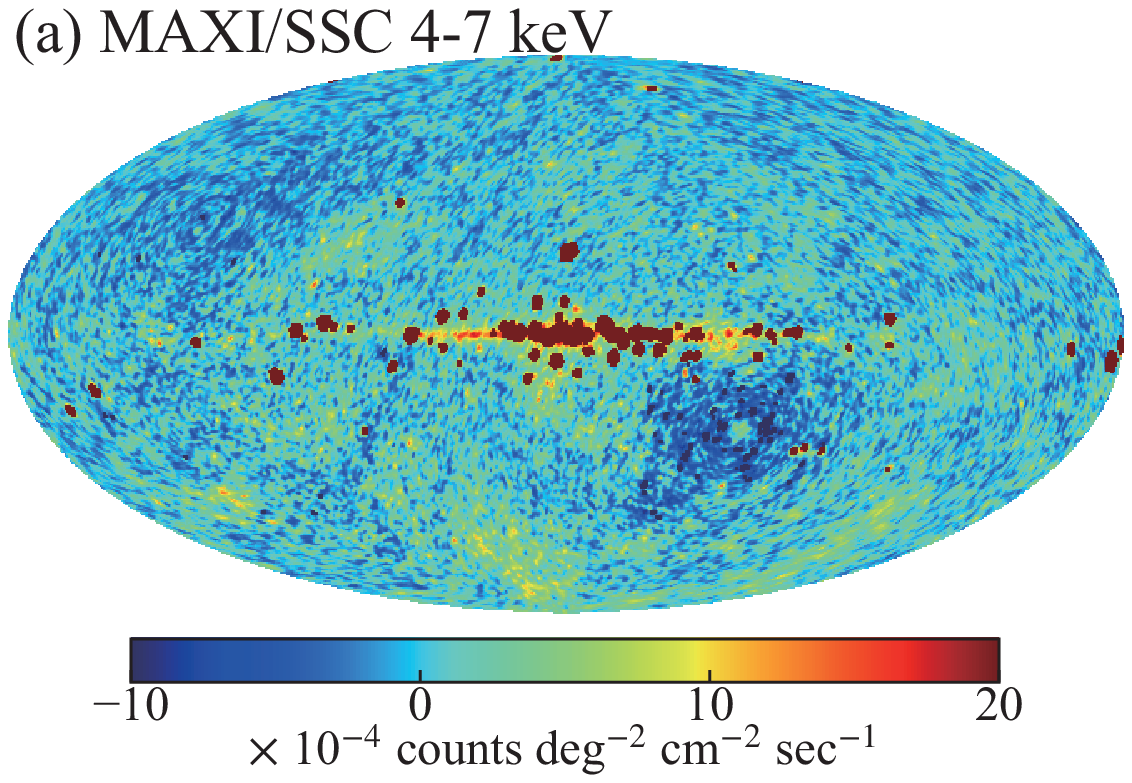}
    \FigureFile(80mm, 80mm){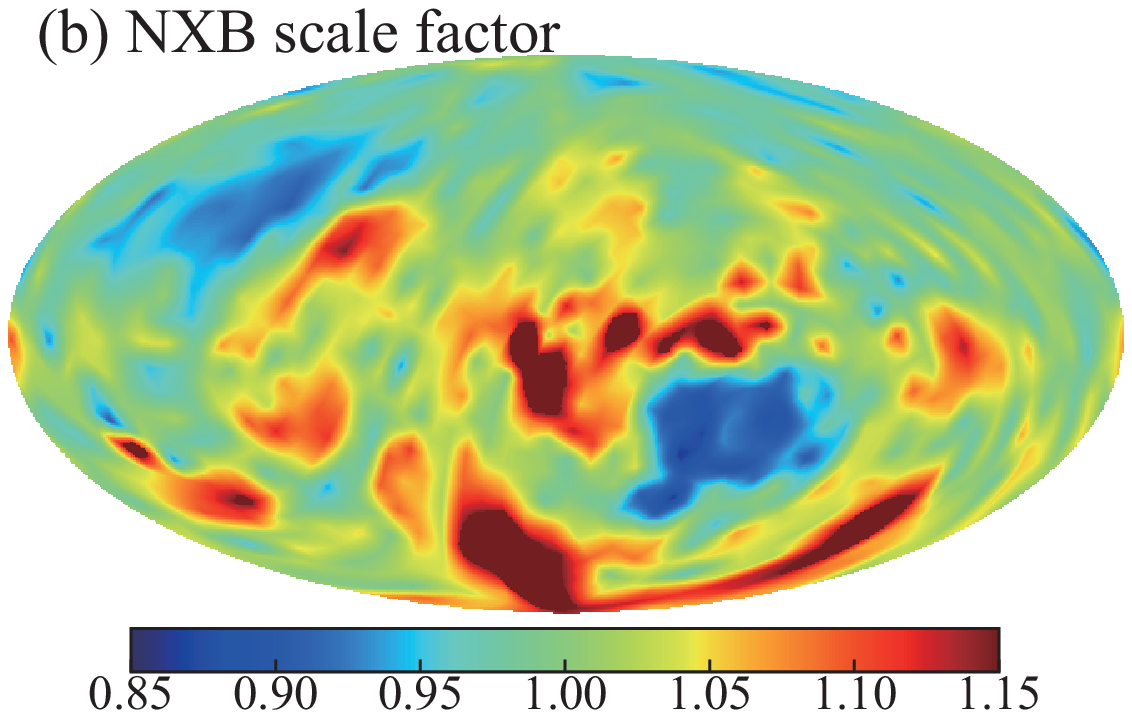}
    \FigureFile(80mm, 80mm){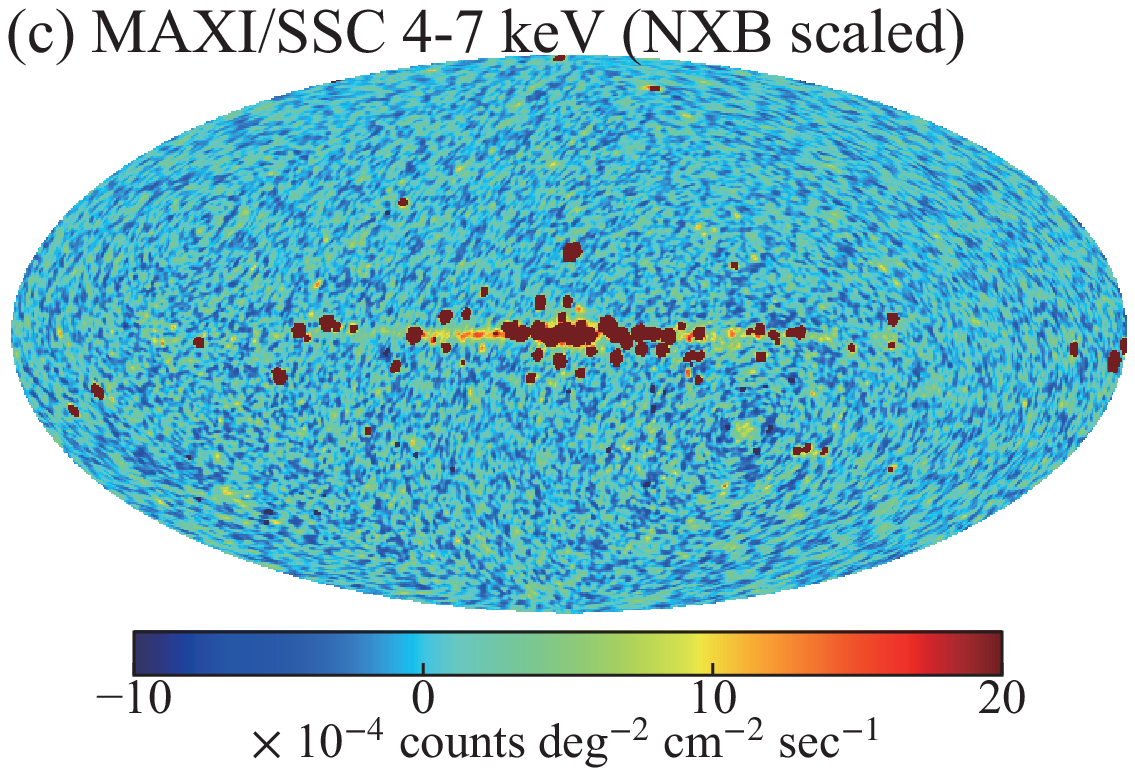}
 \end{center}
 \caption{
 (a) Background (NXB, CXB and GH) subtracted SSC all-sky image 
 in the 4--7\,keV range.
 (b) A sky distribution of the relative NXB intensity,
 to be used in re-scaling the NXB spectral map.
 (c) A 4--7 keV SSC map, obtained by subtracting 
 the scaled NXB spectral map $\Nsie$,
 instead of subtracting $\Nie$ which was used in panel (a).
}
\label{nxbscale}
\end{figure}

To evaluate the reliability of $\Nie$ constructed in this way,
we selected the  4-7\,keV range,
where the CXB and NXB contributions are significant, 
while the GH is negligible because of its low temperature. 
Furthermore, we expect that the CXB and NXB 
are independent of the sky coordinate
except the Galactic absorption of the CXB.
Thus, we have subtracted $\Nie$ from 
the 4--7 keV SSC all-sky image,
and obtained figure\,\ref{nxbscale}(a).
As expected, the result clearly reveals point sources 
and the Galactic-ridge emission.
In addition, we observe several large-scale structures;
one appears as a deficit towards the south-west of the GC,
and another is at its antipodal direction.  
These brightness variations amount to $15\%$ of the NXB.
Because the above two deficit regions are close 
to the poles of the equatorial coordinate,
the structure is considered to be artificial rather than real.
  
Since the SSC scans the sky and accumulates the data for a long time, 
we assumed so far that the NXB time variations 
are almost washed out in our 2-year all-sky maps.
However, the ISS orbit is gradually precessing with a period of 70\,days,
and this may have introduced systematic errors in the NXB estimation.
Namely, the strong NXB variability, correlated with the ISS orbital phase,
coupled systematically with the way of the sky sampling,
to produce the artificial non-uniformity in the NXB map $\Nie$.
Actually, this phenomenon is noticed by the MAXI/GSC \citep{2017symm.conf...29S} avove 3 keV.

The artificial large-scale structure noticed in the
background-subtracted 4--7 keV map (figure~\ref{nxbscale}a)
will also affect the background subtraction in the lower energies
where the EXS should be observed.
However, the shape of the NXB spectrum of the SSC is kept relatively 
unchanged as its intensity varies. 
Therefore, to remove this large-scale artifact,
we have decided to re-scale $\Nie$,
so that the 4--7 keV NXB intensity becomes flat across the sky.
In the normalization process, 
we divide the sky into 768 pixels of about $8^\circ \times 8^\circ$,
removed bright point sources and the Galactic ridge emission,
integrated $\Nie$ at each pixel over the 4--7 keV range,
and divided it by the all-sky average.
These ratios, called the NXB scale factors,
are shown in figure~\ref{nxbscale}(b),
after applying bi-linear interpolation of four nearest neighbours.
Then, by dividing $\Nie$ by this scale factor,
we have derived ``scaled'' NXB spectral map $\Nsie$.
By subtracting this $\Nsie$ instead of $\Nie$,
we finally obtain an improved background-subtracted 4-7\,keV map 
shown in figure\,\ref{nxbscale}(c), 
where we no longer see the large-scale structure.
Likewise, we do not need to rescale the CXB component,
because it is not time variable, and highly uniform all over the sky.

\section{All-sky Map and X-ray Energy spectrum}\label{sec:map_spec}

\subsection{SSC All-sky map}\label{subsec:map}

\begin{figure*}[ht]
\centering
    \FigureFile(80mm,80mm){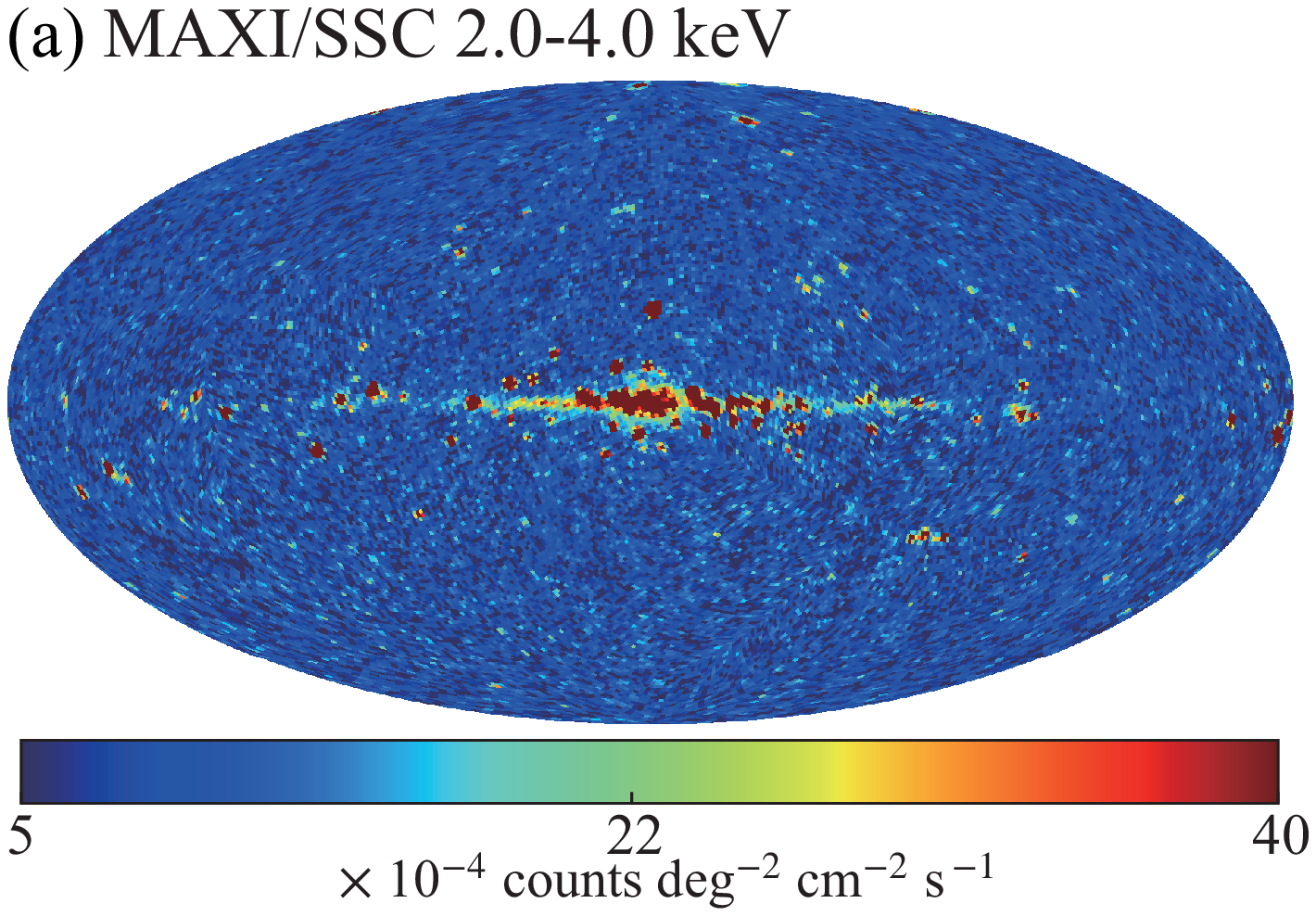} 
    \FigureFile(80mm,80mm){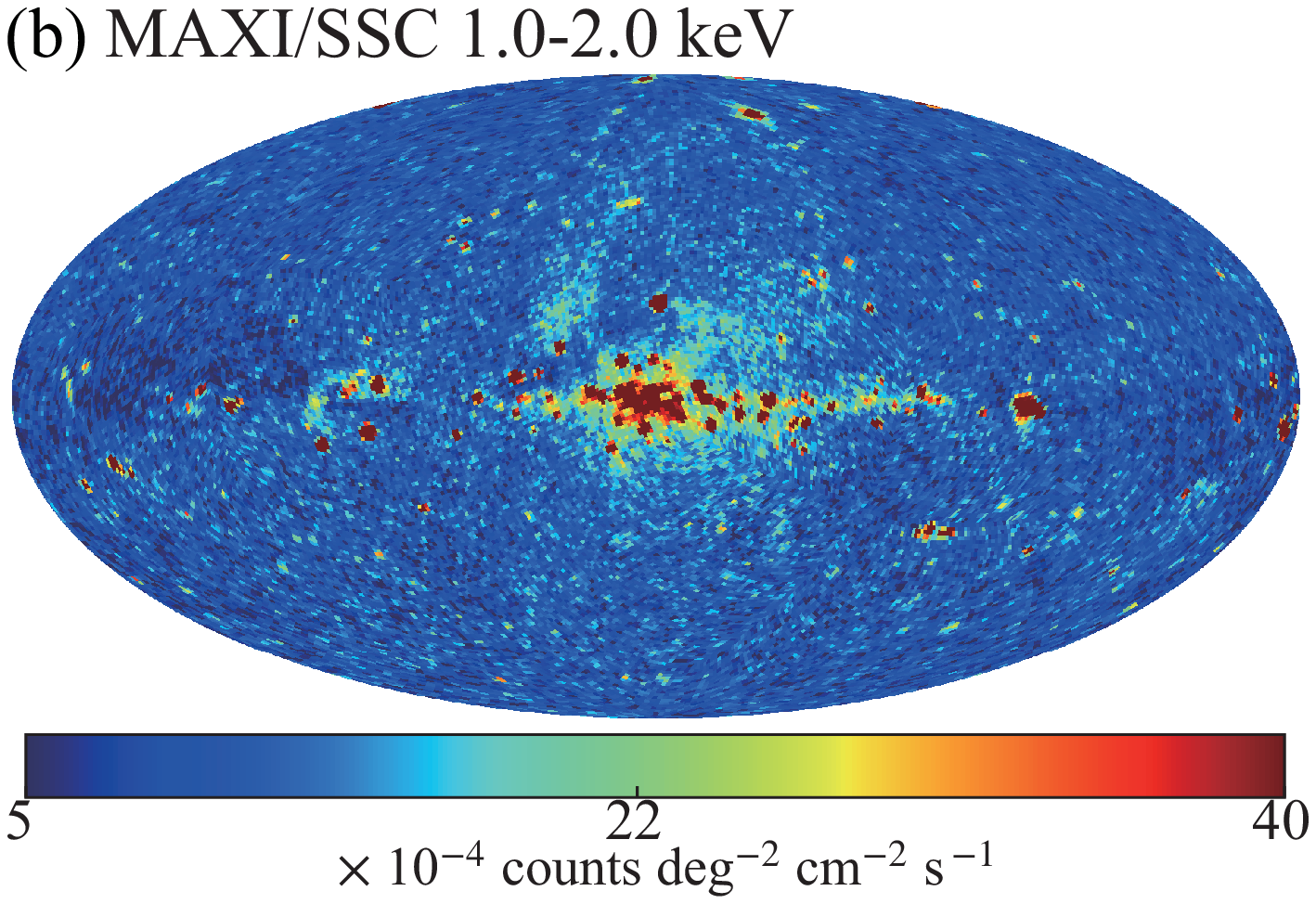} \\
    \FigureFile(80mm,80mm){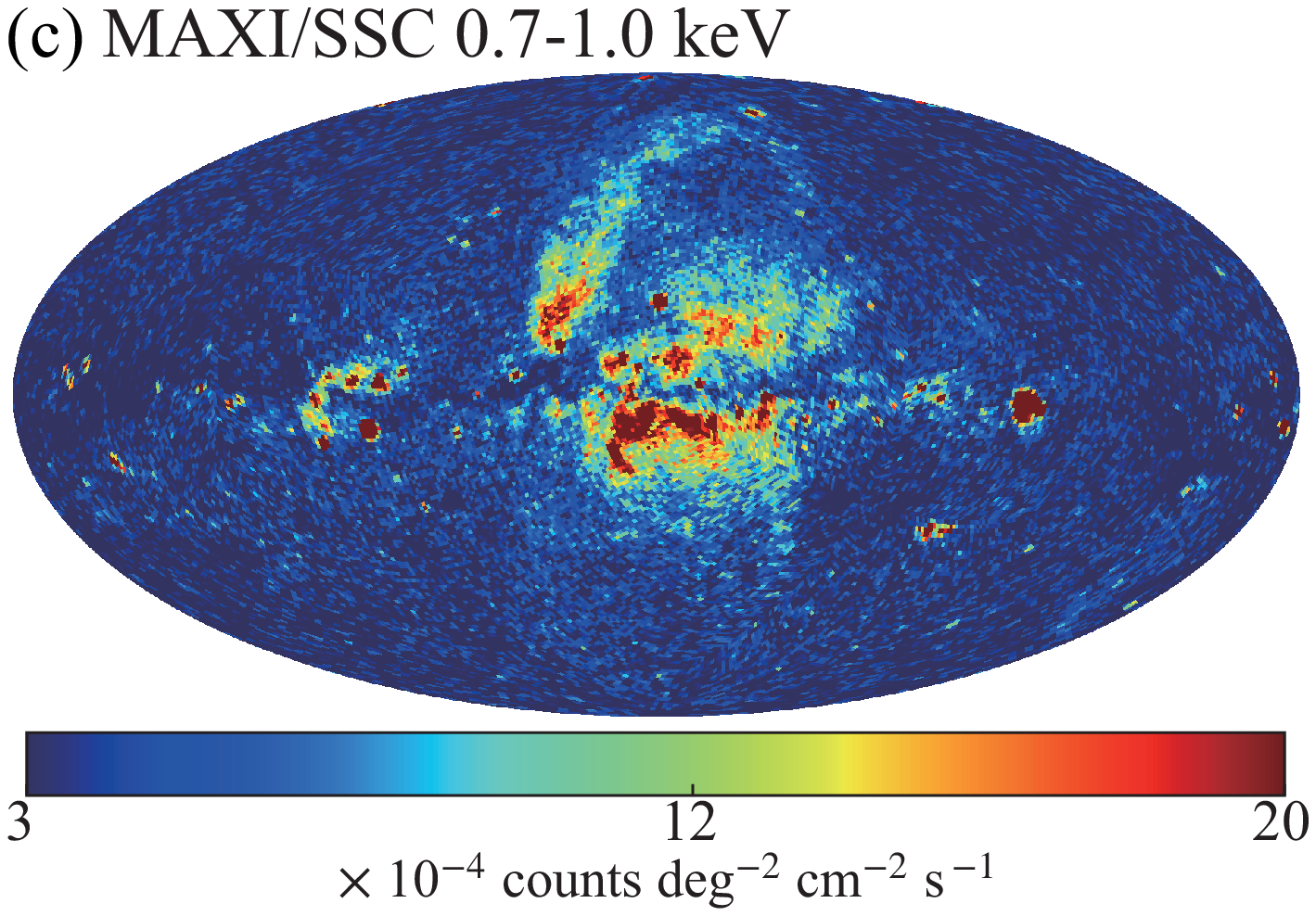}
    \FigureFile(80mm,80mm){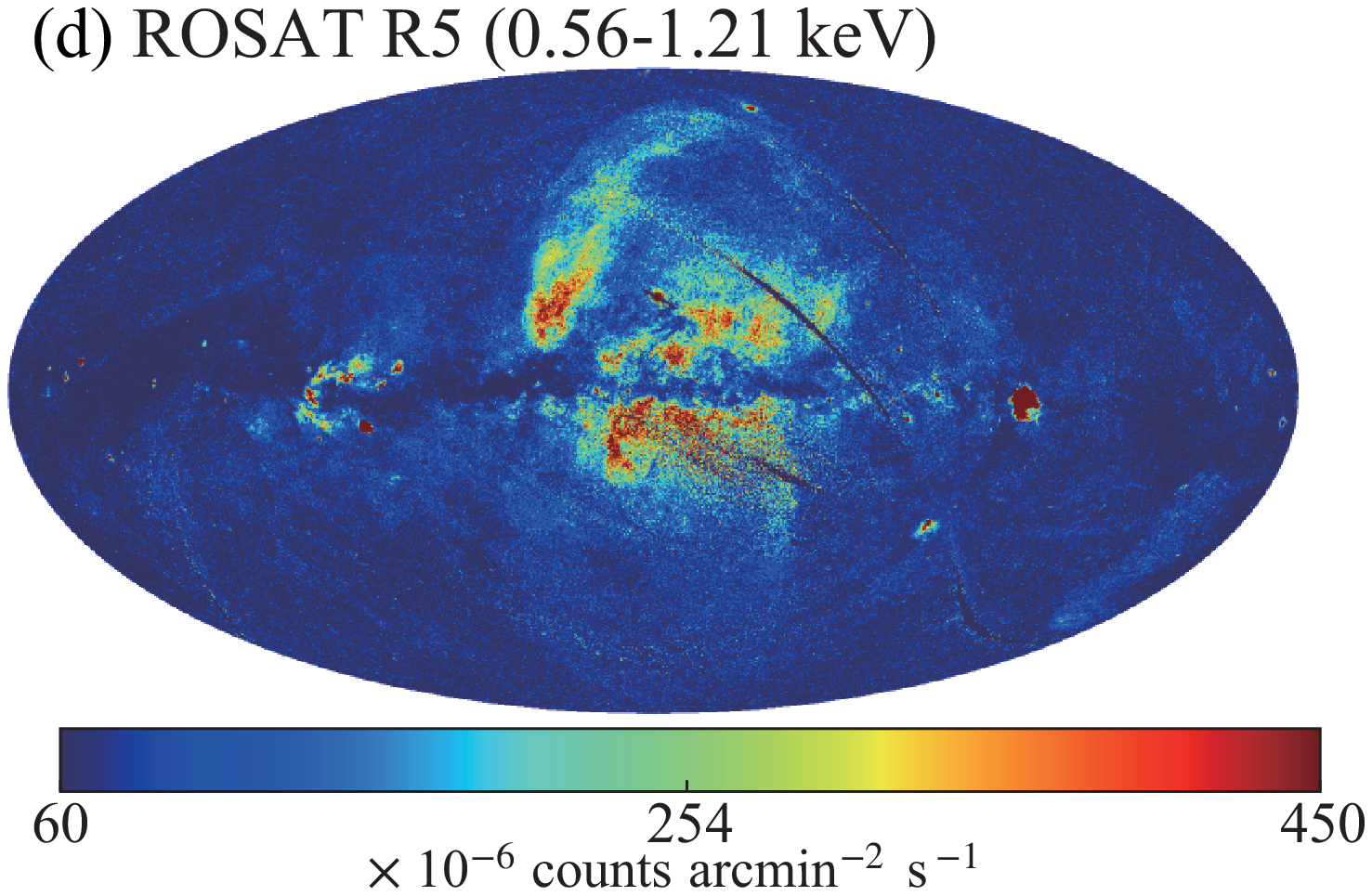} \\
    \FigureFile(80mm,80mm){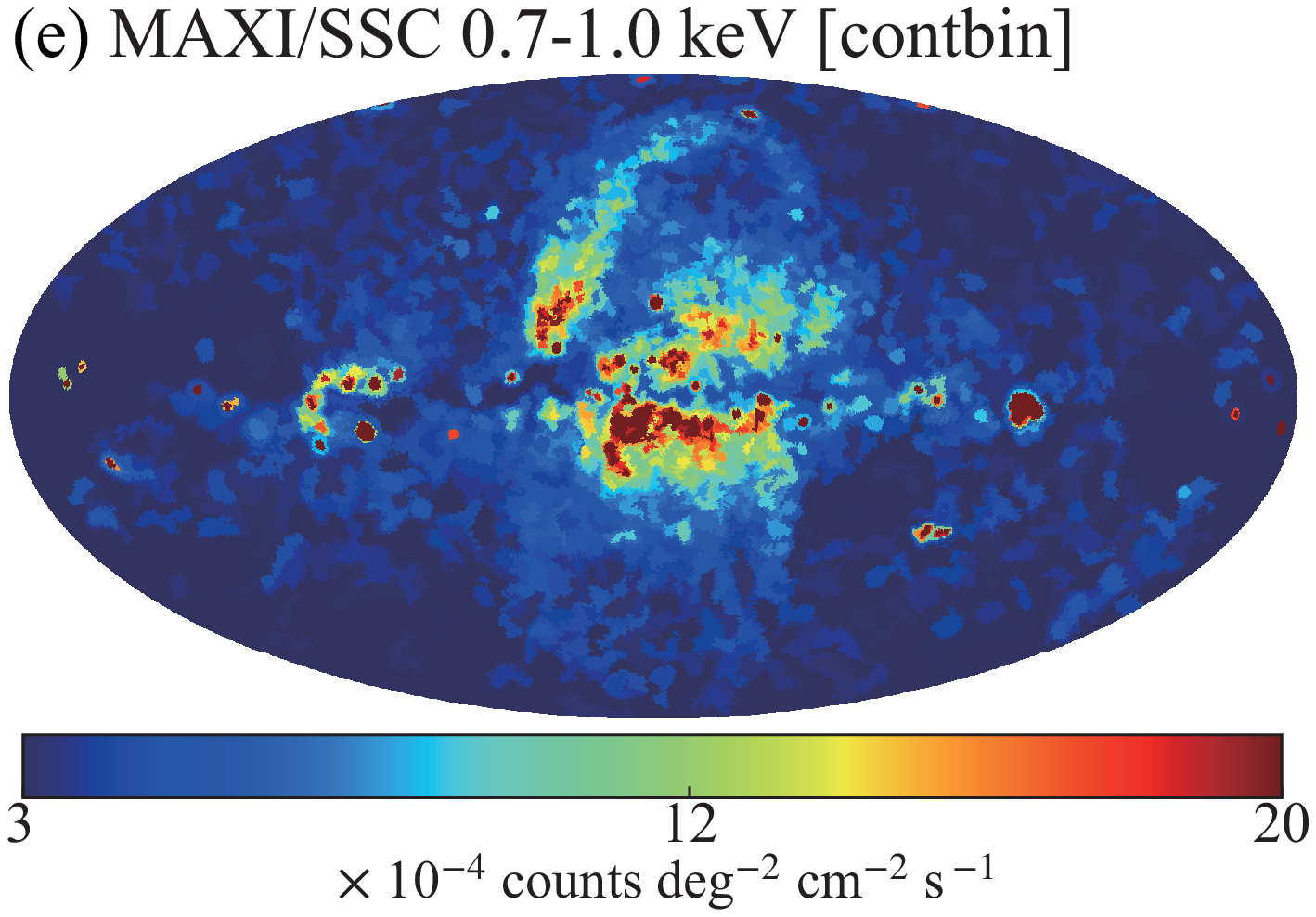}
	\FigureFile(80mm,80mm){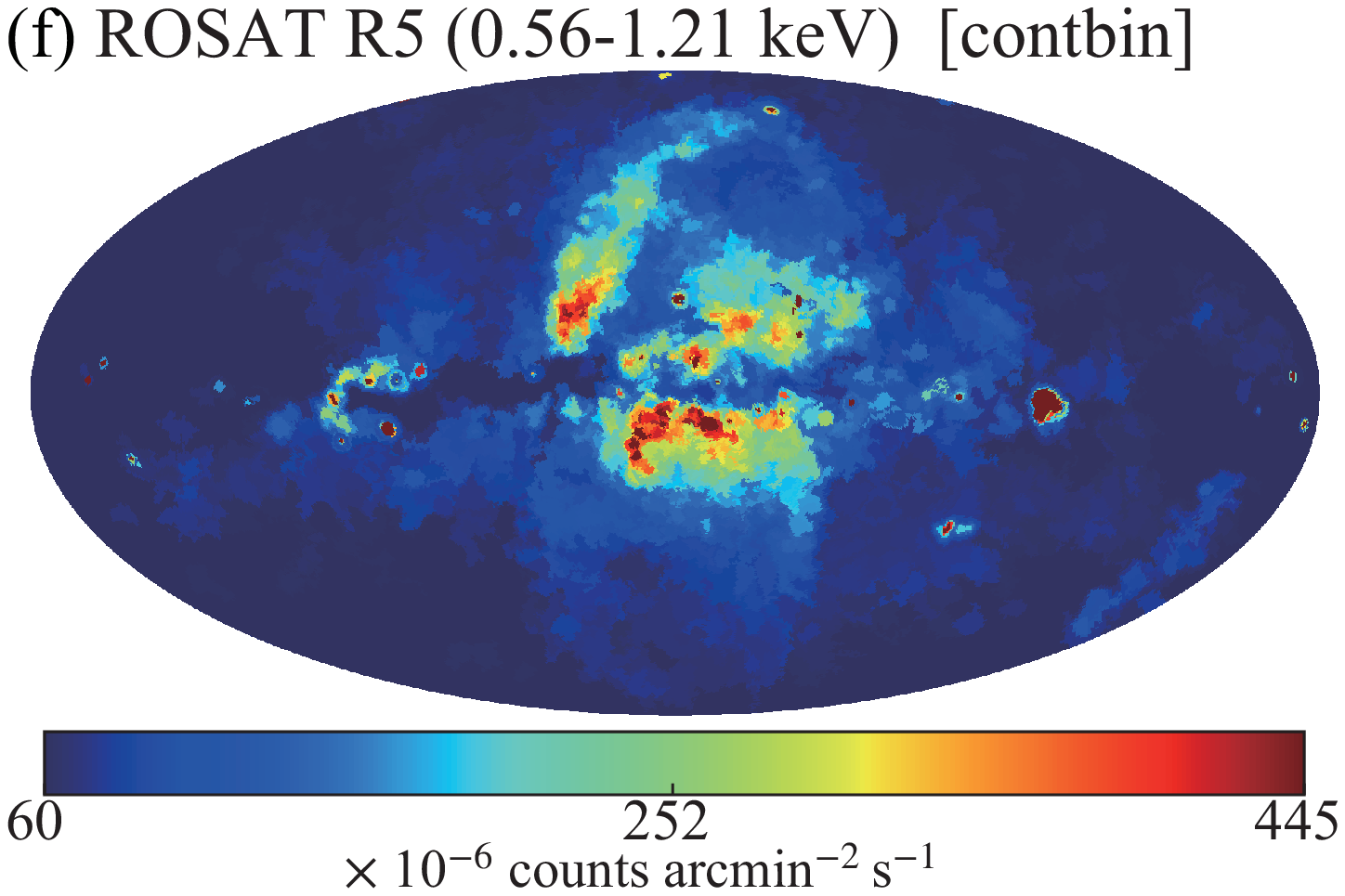}
\caption{
The final MAXI/SSC all sky maps,
in energies of 2.0--4.0\,keV (panel a), 
1.0--2.0\,keV (panel b),
and  \Lend--1.0\,keV (panel c).
The NXB has been subtracted,
but the CXB is inclusive (see text). 
The ROSAT R5 band image is also shown in (d) 
for comparison, with uniform pixel binning. 
Panel (e) shows the same image as (c), but binned 
using the {\tt contbin} method (see text).
Panel (e) is the binned image of (d)
to which the same grouping map as in (e) is applied.
}
\label{SSC_band_map}
\end{figure*}

Through the background subtraction procedure described in the previous section, 
we have generated all-sky maps in the 2.0--4.0\,keV, 1.0--2.0\,keV, 
and \Lend--1.0\,keV energy bands, 
and show them in panels (a), (b), and (c) of  figure\,\ref{SSC_band_map}, respectively.
They are binned in a uniform manner by {\tt HEALPix}. 
In addition, the CXB and the GH, which has once been removed, are restored here,
in order for them to be directly compared with the ROSAT images
which include these components.
Figure\,\ref{SSC_band_map}(d) shows a ROSAT image,
in its R5 band (0.56--1.21 keV) which is similar to the lowest SSC band (panel c).
The ROSAT image is subject to several streaks,
where no exposure was available due the scanning mode condition.

The high energy map (panel a) of the MAXI/SSC is dominated by
many point sources along the Galactic plane, and some extra-galactic sources 
which are consistent with those in the point source catalogs produced with 
the MAXI/GSC \citep{2013ApJS..207...36H} and the SSC \citep{2016PASJ...68S..32T}.  
Except the Galactic ridge emission,
we see no apparent X-ray features that are larger than the SSC PSF.
The medium energy map (panel b) shows similar features to the high energy one,
but it also reveals several extended structures,
including the Cygnus Super Bubble \citep{2013PASJ...65...14K}, 
the Vela Supernova Remnant,
and a large-scale structure apparently extending to north and south
from the Galactic plane within $|l| < 45$ .
This large-scale structure becomes much clearer
in the lowest-energy SSC map (panel c),
and exhibits a very good resemblance to the ROSAT R5 map in panel (d).
Indeed, this is the EXS that we have been aiming to study.

Both in the ROSAT and the SSC images,
the faintest parts of the EXS suffer statistical fluctuations.
Therefore, we employed the {\tt contbin} method by \citet{2006MNRAS.371..829S}, 
which allows us to perform variable-size binning
to achieve similar signal-to-noise ratios.  
The binning process was conducted using the {\tt contbin} tool, 
with its input parameters of {\tt sn}=10 and {\tt smoothsn}=5,
where the \Lend-1.0\,keV raw SSC image, its corresponding NXB image, 
and the exposure maps were also given as the inputs.
Then, at each bin of the smoothed map to be produced,
the surface brightness was calculated in the usual way,
by subtracting the NXB map from the total map, 
and dividing it with the exposure map,
each summed over the same binned area.
In this  way, the lowest-band SSC image in figure\,\ref{SSC_band_map}(c) 
has been converted to figure\,\ref{SSC_band_map}(e).
Using the same binning map as used to smooth the
\Lend-1.0\,keV SSC image, 
the ROSAT image in figure\,\ref{SSC_band_map}(d) 
was converted into figure\,\ref{SSC_band_map}(f) 
where the streaks were mostly swept out. 

\begin{figure*}
 \begin{center}
      \FigureFile(150mm, 150mm){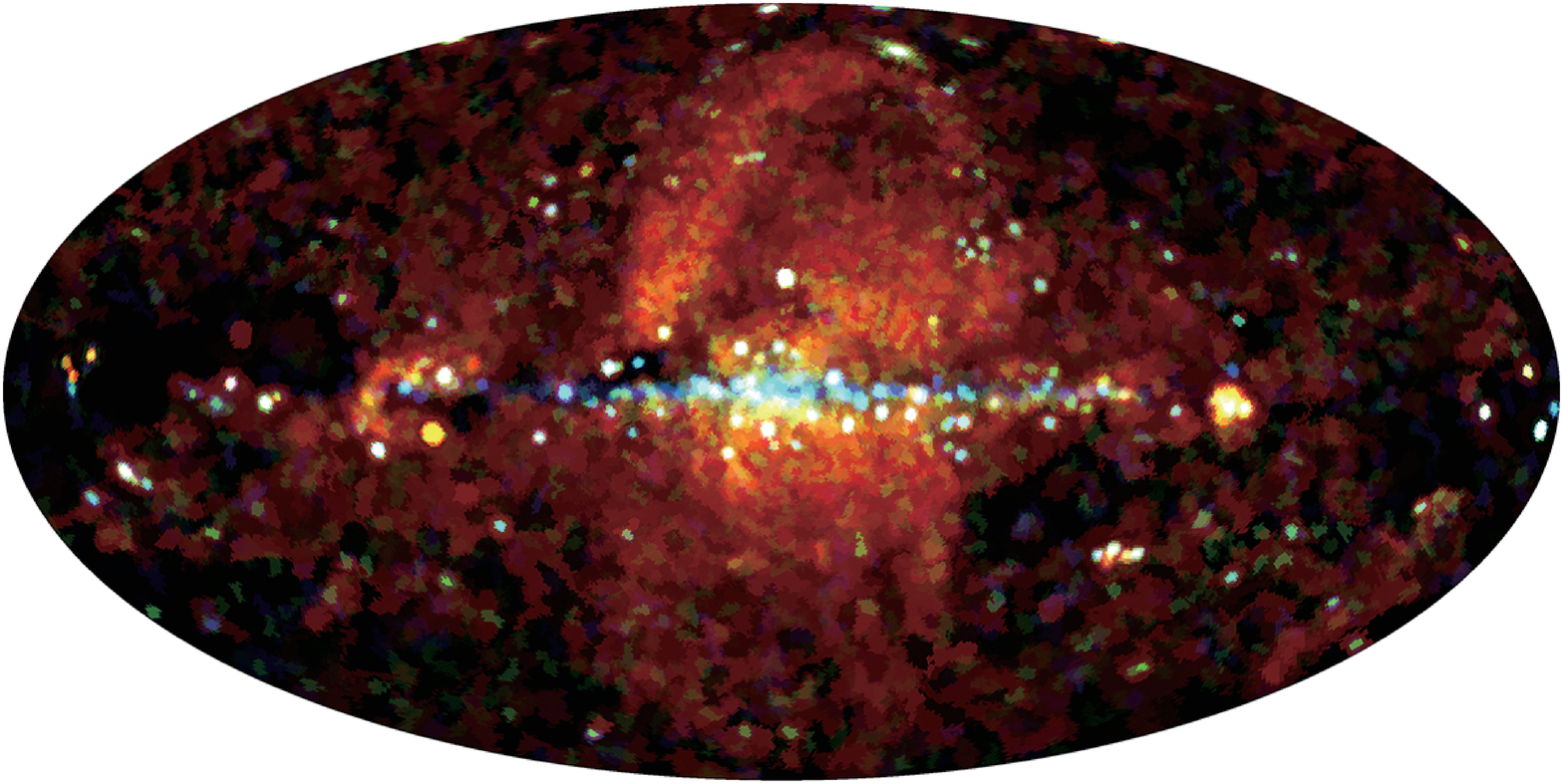}
  \end{center}
  \caption{A pseudo-color all-sky map by the SSC. 
  Blue, green and red represent brightness
  in the  2.0--4.0\,keV, 1.0--2.0\,keV and \Lend--1.0\,keV
  energy bands, respectively.}
  \label{rgbmap}
\end{figure*}

Finally, by combining the three-band SSC maps,
we have created a pseudo-color map as shown in figure\,\ref{rgbmap}.
There, blue, green and red represent the high, medium and low energy X-rays.

\subsection{Comparison with the ROSAT all sky survey}
\label{subsec:comparison}

The SSC maps presented above are the first all-sky soft X-ray images
since the ROSAT all sky survey was done in 1990-1991 \citep{1999hxra.conf..278F}.  
Indeed, the smoothed \Lend-1.0\,keV SSC sky map (figure\,\ref{SSC_band_map}e)
is amazingly similar to that with ROSAT (figure\,\ref{SSC_band_map}f),
in spite of the small difference in the energy band
(\Lend-1.0\,keV vs.0.56-1.21\,keV),
the much better energy resolution of the SSC,
the much higher angular resolution of ROSAT,
and above all, the 20 years of time difference between the two surveys.
It is hence confirmed that the two missions have caught
essentially the same celestial phenomenon.

\begin{figure}[ht]
  \begin{center}
       \FigureFile(80mm,80mm){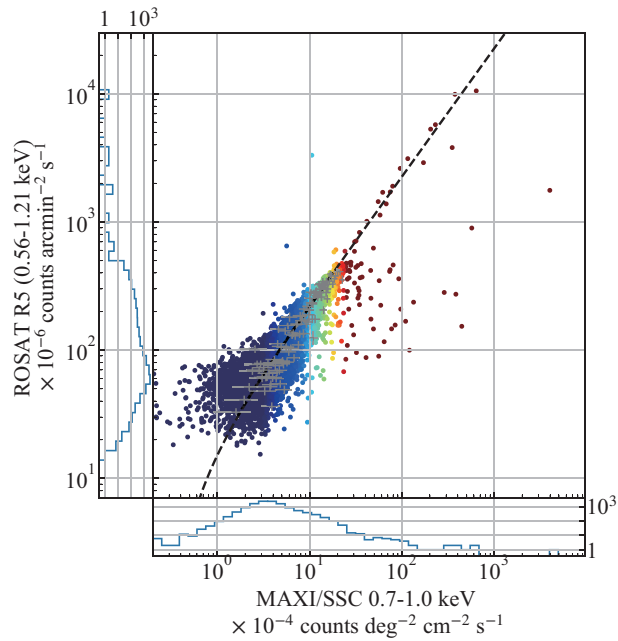}
  \end{center}
\caption{
A scatter plot between the SSC brightness and that of ROSAT,
in figure\,\ref{SSC_band_map} (c) and (d), respectively,
where the pixel resolution is converted to $\nside$=32.  
The data points are colored referring 
to the color bar of figure\,\ref{SSC_band_map}(e).
Typical error bars are also shown in gray.  
The best fit linear relation is shown in dashed line (see text).
}
\label{SSC_ROSAT}
\end{figure}

To quantitatively compare the soft X-ray skies between ROSAT and the MAXI/SSC,
we divided the entire sky into 12288 pixels
using $\nside$=16 of the HEALPix coordinate.
Figure\,\ref{SSC_ROSAT} is a scatter plot between the SSC and ROSAT 
brightness values measured at each of the 12288 pixels. 
There, the brightest pixels above $10^3$ (in the ROSAT scale)
come from the EXS, the Vela Supernova Remnant, and the Cygnus Super Bubble.  
Including these bright pixels, the two quantities show a very good correlation.
However, due to the much larger PSF of the SSC, 
some pixels near bright point sources become much  
higher in the SSC than in the ROSAT map;
these points are distributed toward the lower right region in figure\,\ref{SSC_ROSAT}.

After eliminating the pixels near bright point sources, 
we fit the data (10532 pixels) by a linear least squares procedure
considering both MAXI/SSC and ROSAT \citep{1997ApJ...485..125S} statistical errors.
The best fit, shown in a dashed line in the figure,
is expressed as 
\[
y = 22.8 \left\{ x -(7.6 \pm 0.4) \right\}
\]
where $x$ and $y$ denote the SSC and ROSAT measurements, respectively.
Although the result implies a small offset,
the correlation is seen to hold over 3 orders of magnitude in brightness.

\begin{figure}[ht]
  \begin{center}
       \FigureFile(80mm,80mm){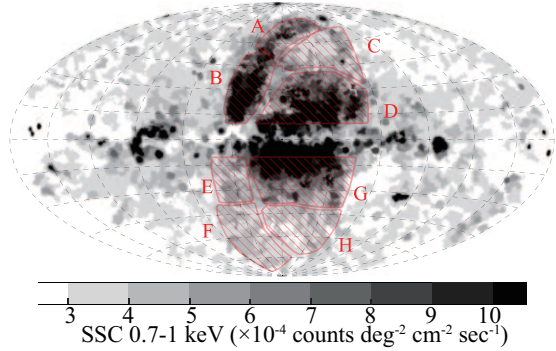}
  \end{center}
\caption{A gray-scaled image of figure\ref{SSC_band_map} (e) 
where Areas A to H are indicated by hatching.  
The spectral analysis is carried out for each area,
and the results are summarized in table\,\ref{tab:spec_phase}.}
\label{regmap}
\end{figure}

\subsection{Spectral analysis of bright regions}
\label{subsec:specana}

\begin{figure*}[ht]
  \begin{center}
       \FigureFile(50mm,50mm){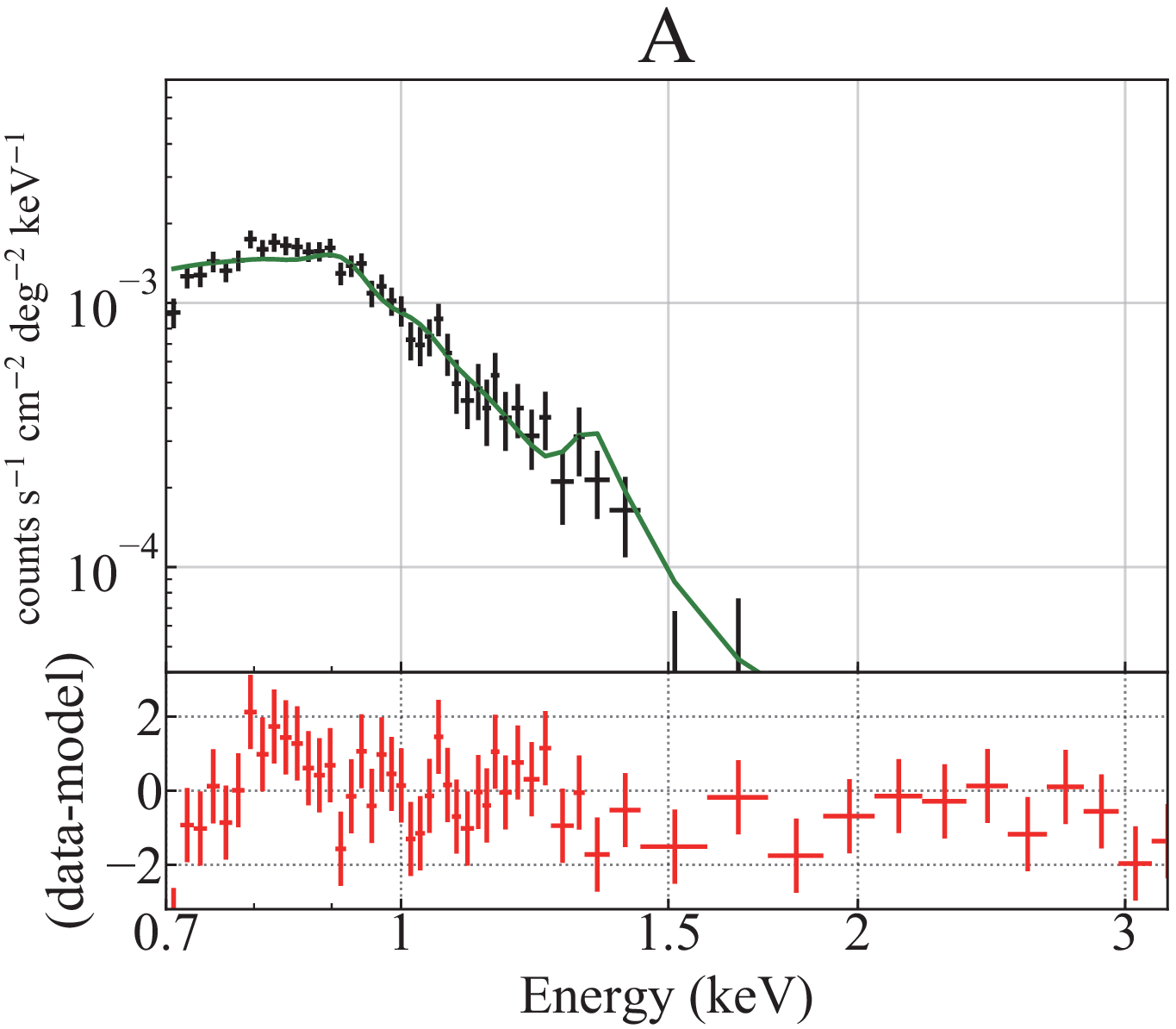}
       \FigureFile(50mm,50mm){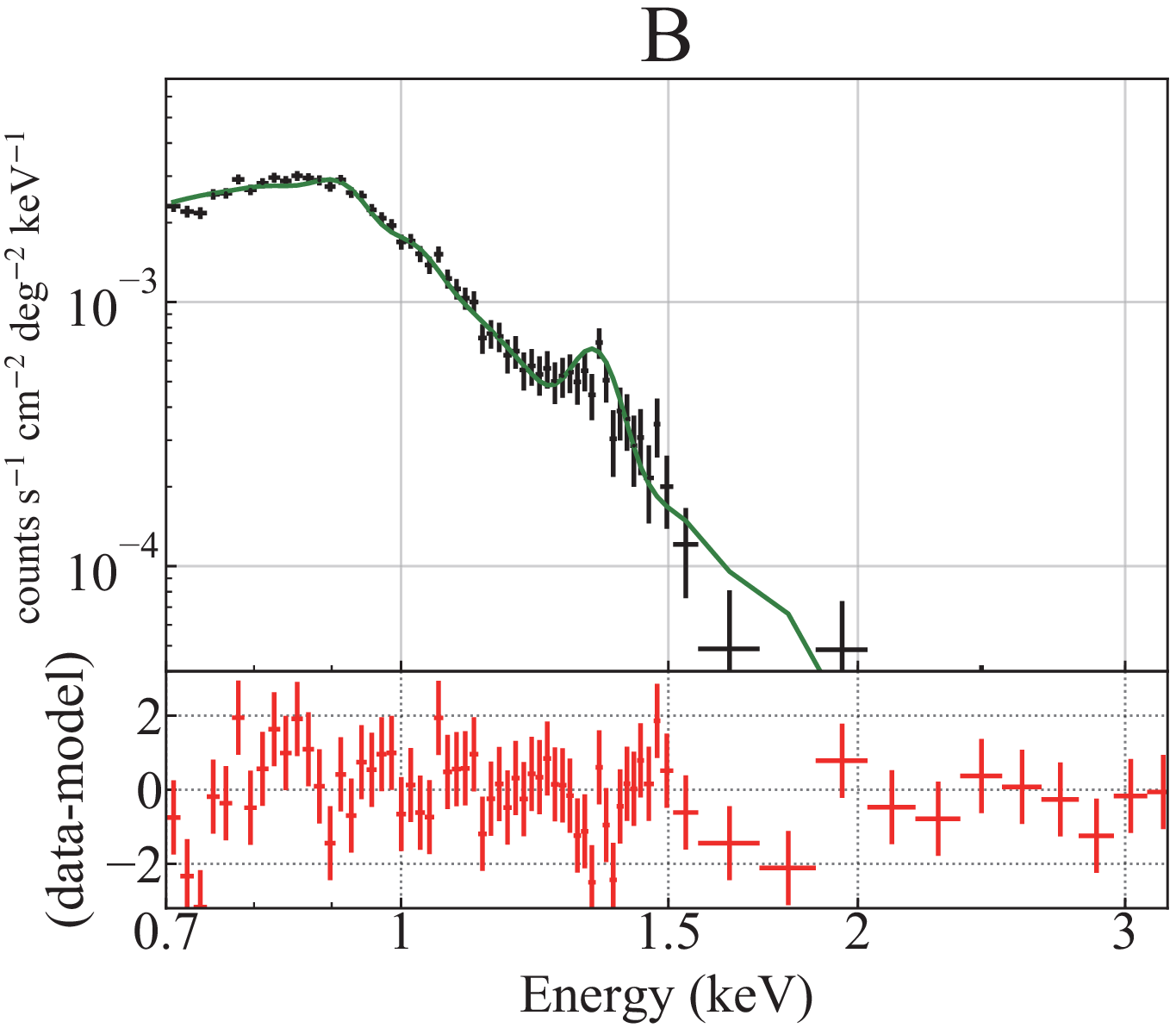} 
       \FigureFile(50mm,50mm){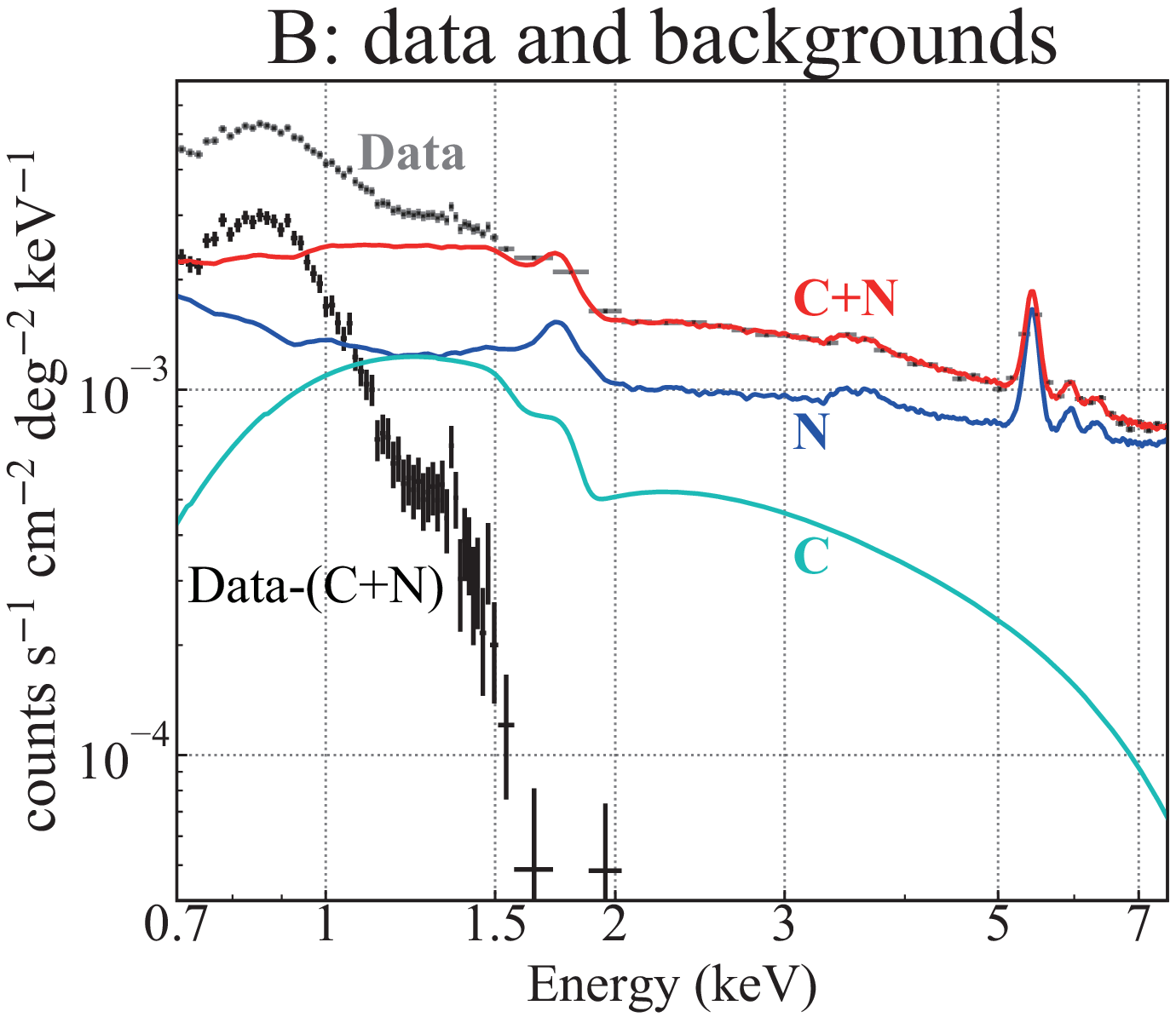}\\
       \FigureFile(50mm,50mm){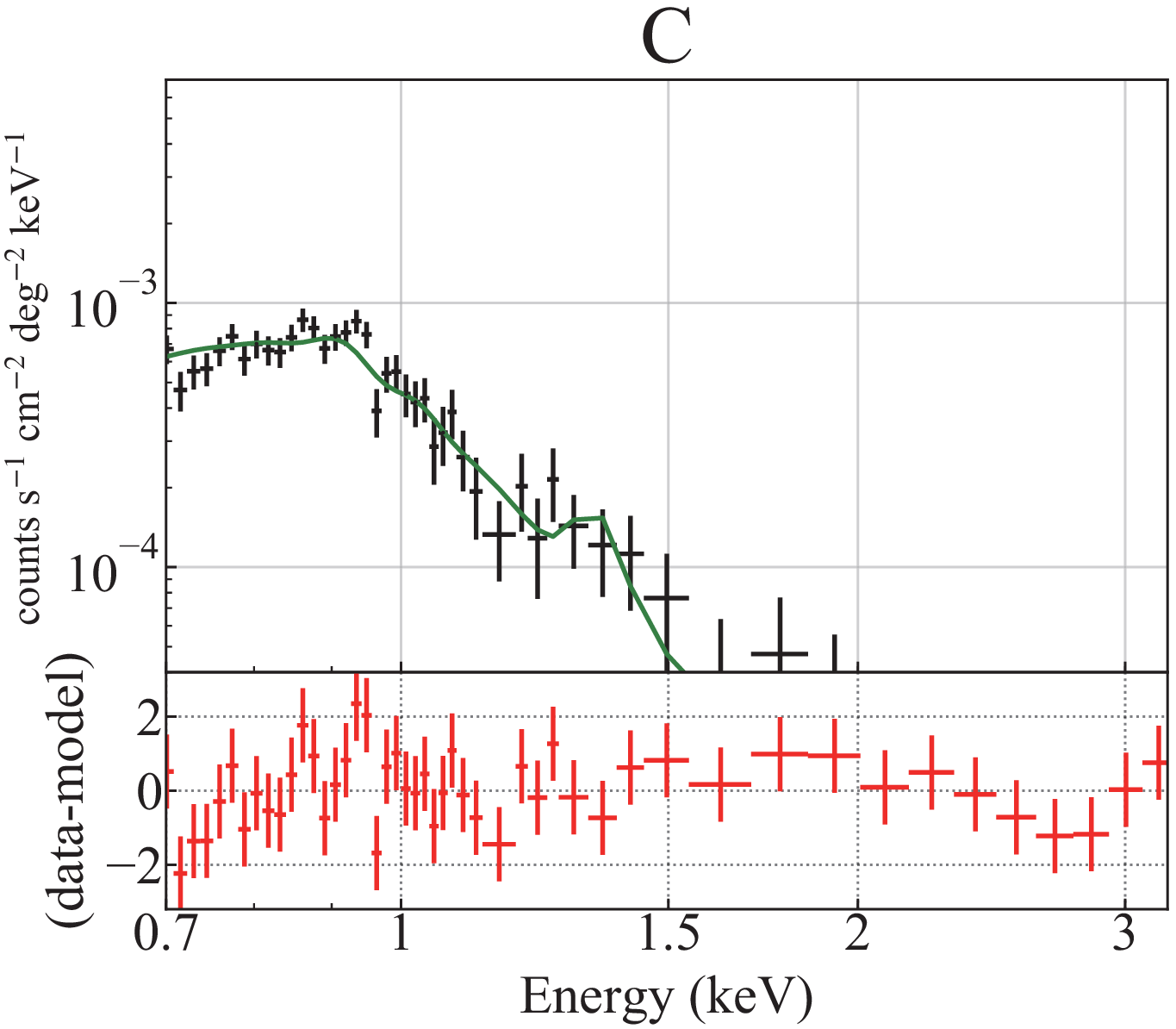}
       \FigureFile(50mm,50mm){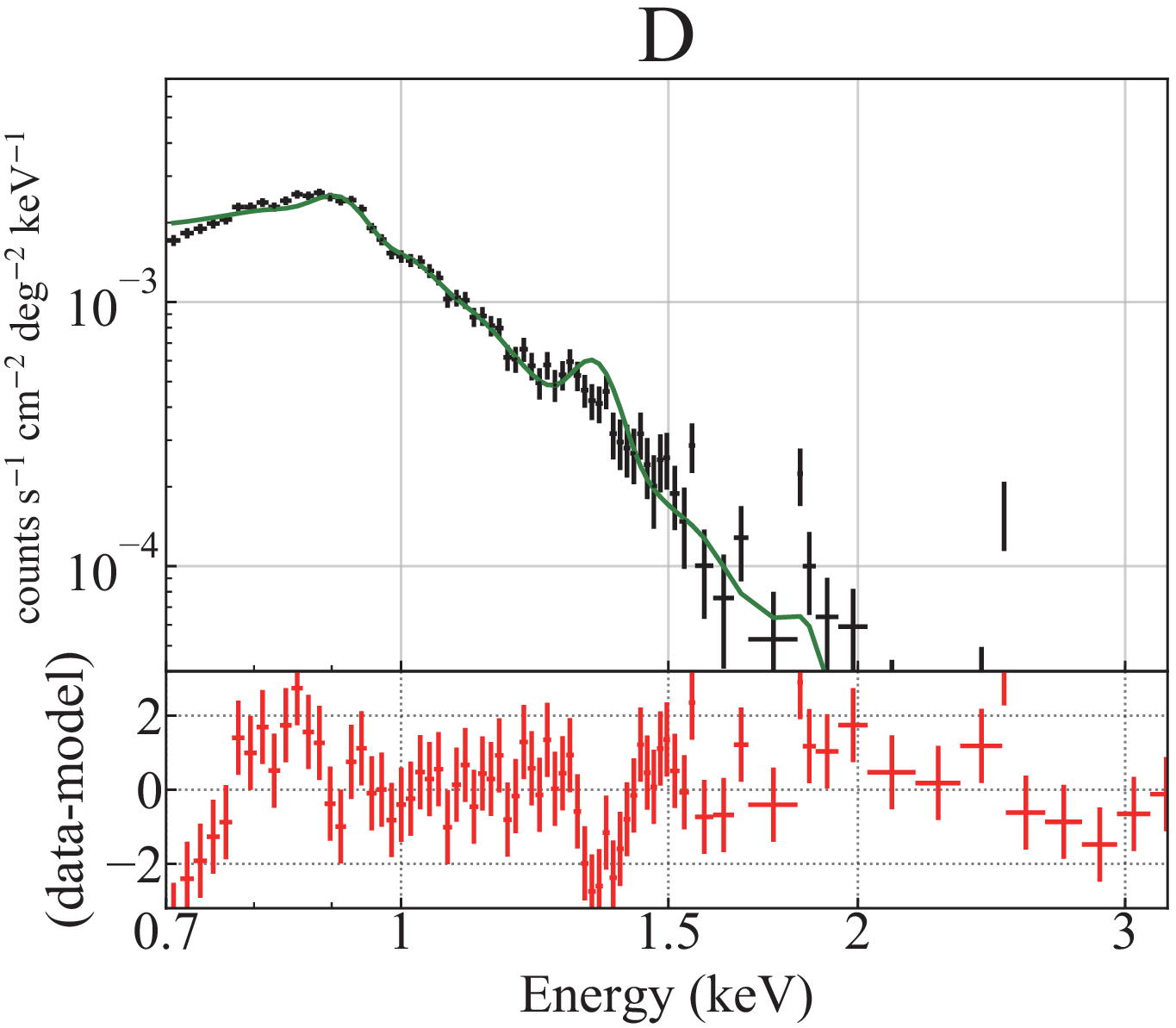}
       \FigureFile(50mm,50mm){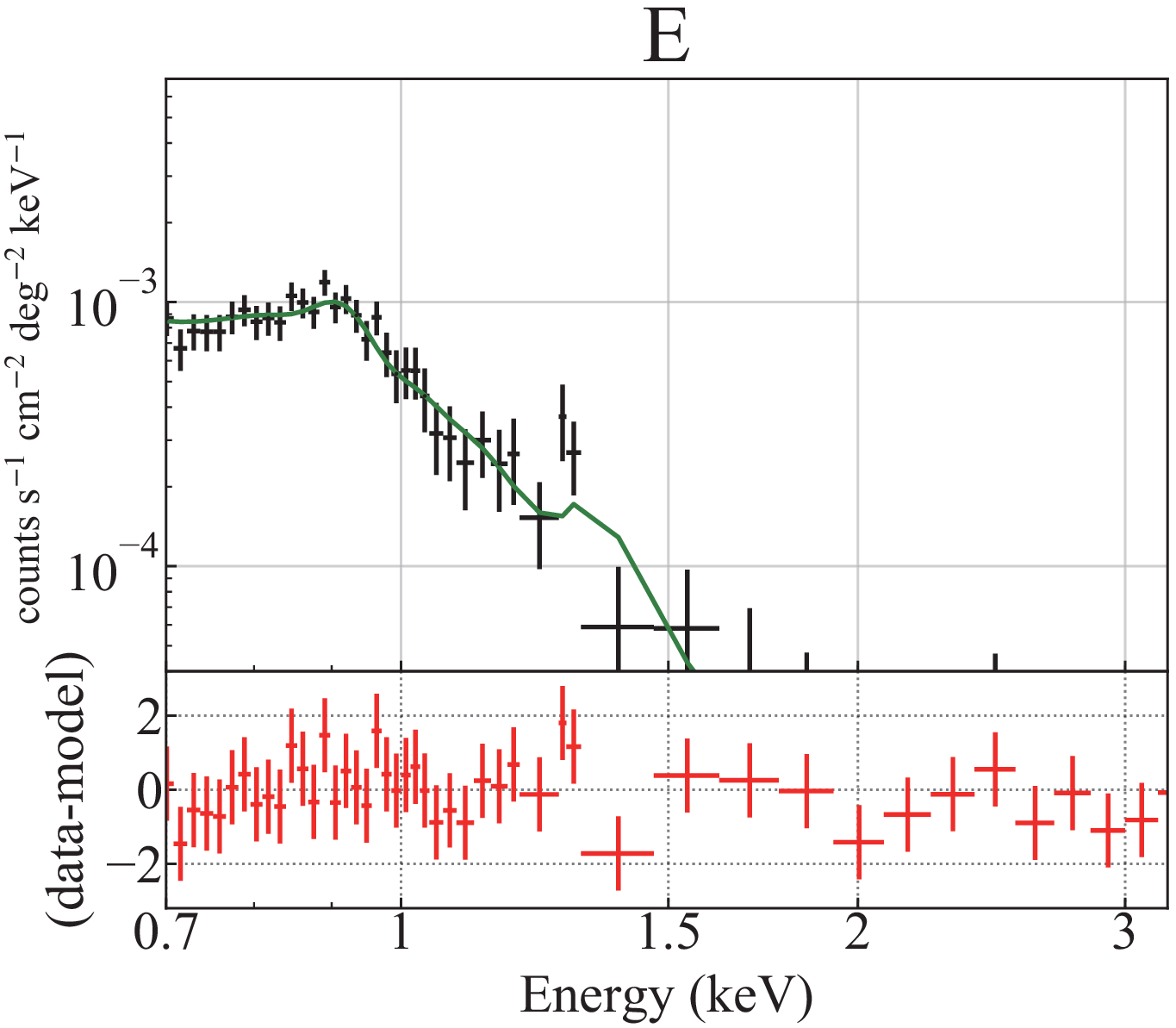} \\
       \FigureFile(50mm,50mm){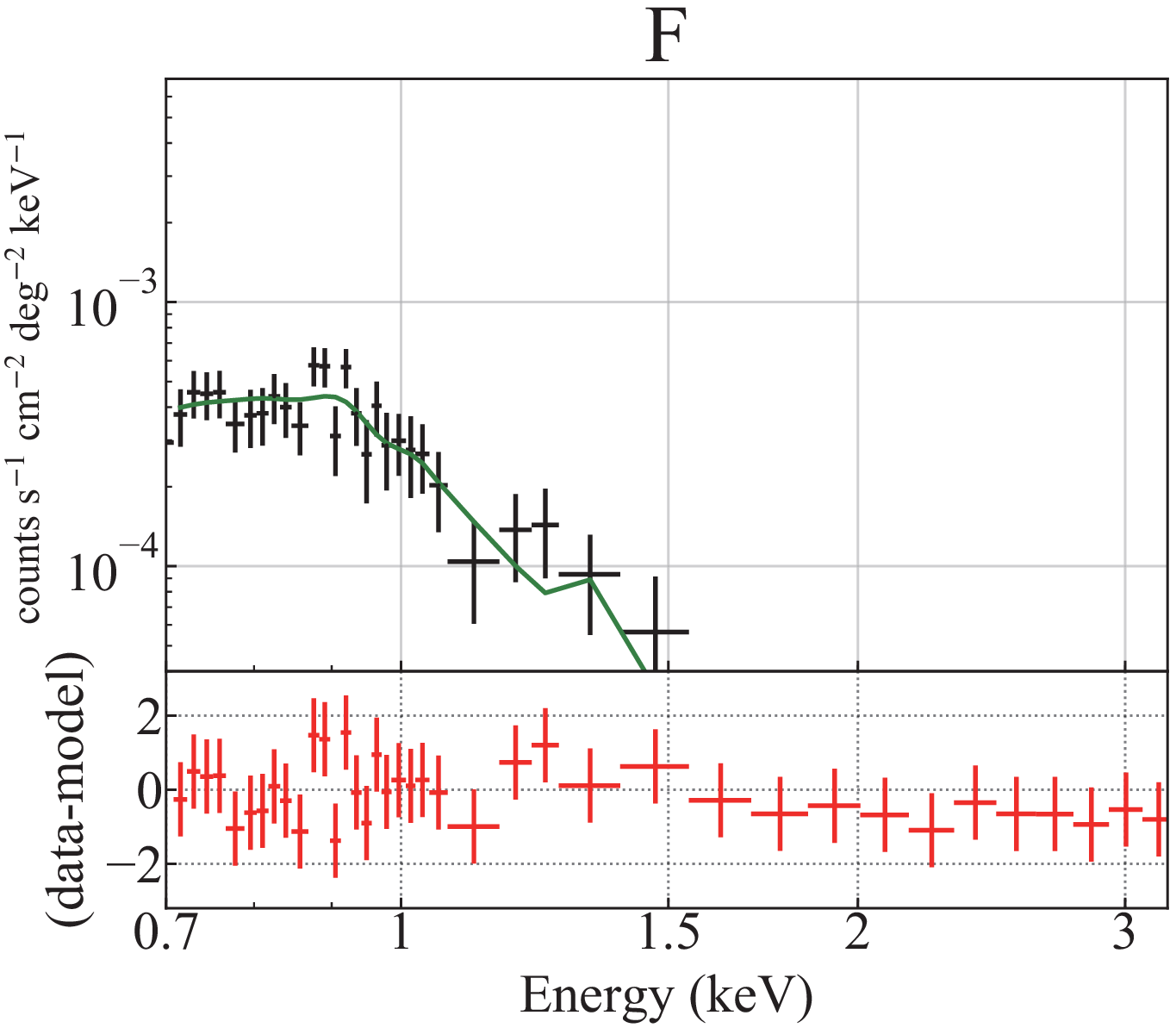}
       \FigureFile(50mm,50mm){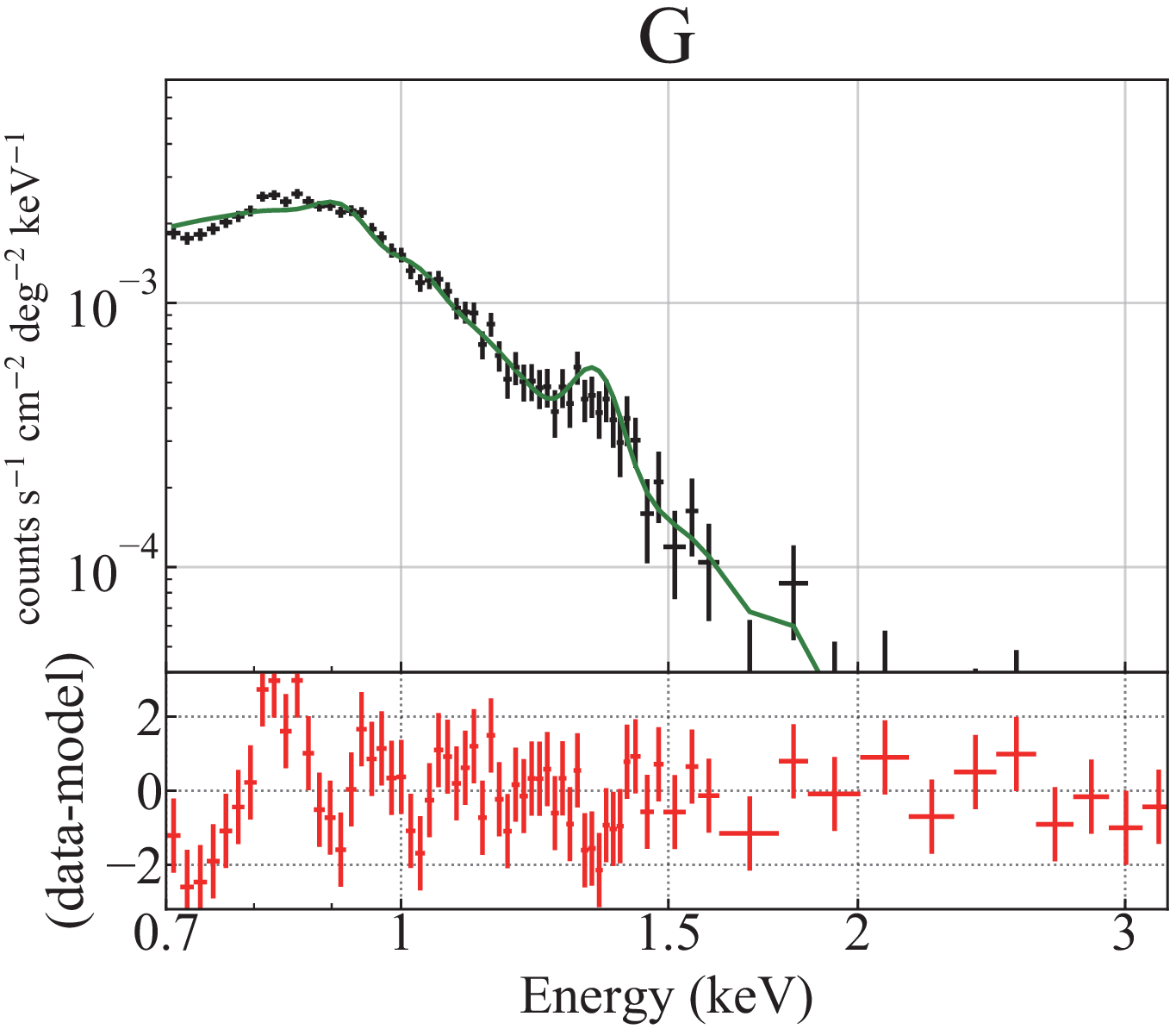}
       \FigureFile(50mm,50mm){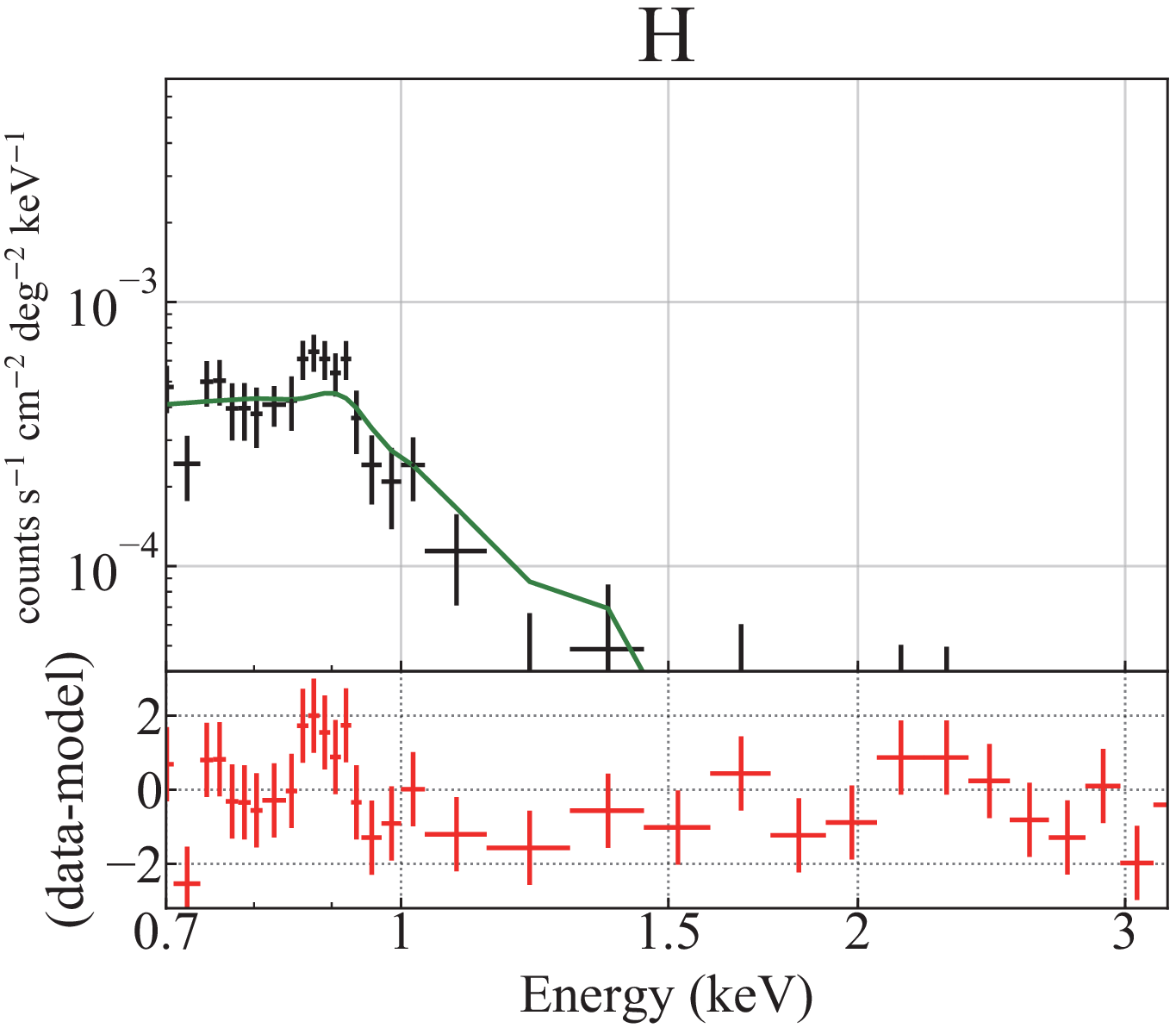}
  \end{center}
\caption{
The SSC spectra derived from the eight Areas 
defined in figure~\ref{regmap}, labeled as A through H.
Black are the background-subtracted  data, 
a green solid line shows the best-fit model.
and red points are fit residuals.
The top-right panel shows the spectral composition for Area B,
where the raw on-source data are in gray,
the CXB and the NXB are in cyan and blue, respectively,
and their sum is in red. 
The black is the spectrum obtained by subtracting the NXB and CXB.
}
\label{spec_fig}
\end{figure*}

To study the SSC spectra of the EXS, 
we divided its apparently bright part at $|\ell| \lesssim 60^{\circ}$ 
into eight areas, from Area A to Area H, 
as shown in figure\,\ref{regmap}. 
The background-subtracted SSC spectra,
derived from these 8 areas, are presented in figure~\ref{spec_fig}.
Thus, the eight spectra are relatively similar.
They all exhibit an emission feature at $\sim 1.4$ keV,
and a broad hump at $\sim 0.8$ keV,
which are identified with Mg-K lines of He-like or
similar ionization states,
and possibly Fe-L line complex, respectively.
Thus, the spectra are definitely thermal, as already known \citep{2003MNRAS.343..995W}.

We first analyze the spectrum of Area B,
which is the brightest one 
and includes the radio feature known as ``North Polar Spur''.  
The top-right panel of figure\,\ref{spec_fig} 
shows the spectral composition of this Area.
There, the highest gray data points indicate the raw on-source data
before the background subtraction, 
while the CXB and NXB which we subtracted
are shown in cyan and blue, respectively,
with their sum in red.  
The background-subtracted data are in black,
which is identical to the black spectrum
at the top-middle panel.

Before proceeding to the spectral fitting,
we need to construct area-specific X-ray absorption models,
because the eight Areas are individually extended
over hundreds of square degrees.
According to the HI column density map of
the Leiden/Argentine/Bonn (LAB) survey
\citep{2005A&A...440..775K},
the line-of-sight equivalent Hydrogen column density \nh\
scatters by more than an order of magnitude across each Area.
Therefore, a spectrum accumulated over such a region
would be absorbed in a complex way,
of which the effect can no longer be
expressed by a single value of \nh.
To take into account the wide distribution of \nh, 
we created a local absorption model,
one for each Area.
We first calculated the absorption curve for each $\nh$ 
using the {\tt TBabs} model in Xspec,
and averaged these curves over the $\nh$ distribution 
in the target region,
weighted by the \Lend--1.0\,keV photon number.
According to \citet{2013MNRAS.431..394W},
the $\nh$ values from the LAB map were
multiplies by a factor of 1.25.
The area-specific absorption factor as constructed above,
denoted by $A$, was further modified
by another factor $\fgal$ as $A'=A^{\fgal}$,
which was then applied to the spectral model
under a constraint of $\fgal \le 1.0$.
This $\fgal$ can be used as a distance indicator 
of the EXS emitting region within the Galactic disk. 
If $\fgal$ is low, the source is inferred to be at a nearby location,
and if it becomes unity, 
the source is considered to reside outside the Galactic disk. 

The spectrum of Area B was tentatively fitted by a power law model,
using the absorption factor and $\fgal$ as explained above, 
but the fit was unsuccessful with $\chi^2$/d.o.f=446.0/67. 
Then, we applied a thin thermal emission model, 
{\tt vapec} in {\tt Xspec} version 12.10.0c,
with the abundance table by \citet{2000ApJ...542..914W} 
wherein all elements are linked together. 
The fit was much improved, but not yet acceptable ($\chi^2$/d.o.f=234.1/66),
with $\fgal>$0.9, the plasma temperature of $\kt \sim 0.3$ keV,
and the metallicity  of $Z \sim 0.10~\zo$. 
Finally, the metallicity ratio of Fe relative to the others 
($\zfe/Z_{o}$) was set free,
while the elements other than Fe were constrained to have the same abundance.
The fit has become acceptable with $\chi^2$/d.o.f= 79.2/65,
yielding $\fgal>0.6$, $\kt=0.31 \pm 0.01$ keV, 
$Z=0.51_{-0.14}^{+0.1}~\zo$, and $\zfe/Z_{o}=0.52_{-0.06}^{+0.03}$.  
Figure\,\ref{spec_fig}B (top center) shows the data 
with its best fit model (solid line)
as well as the residuals (in the bottom panel).

\begin{table*}
  \caption{The best-fit model parameters 
  to the MAXI/SSC spectra from Areas A to G in figure~\ref{regmap}.} 
  \begin{center}
\begin{tabular}{ c c c c  c c c c c  } \hline
Area ID & Area Size$^{*}$\
&$\fgal^{\dagger}$ & $\nhgal^{\ddagger}$ & $kT^{\S}$  & $Z^{\parallel}$  & $\zfe$/$Z_{o}^{\#}$  & EM$^{{**}}$ & reduced $\chi^2$ (dof) \\ \hline
A & 424.4 & 1.0$^\ddagger$ & 0.18--0.33 & 0.31$\pm0.02$  & 0.30$^\ddagger$ & 0.50$^\ddagger$  & 3.8$\pm0.4$ & 1.22 (53) \\ 
B & 783.1 & 1.0($>$0.6) & 0.35--1.56 & 0.31$\pm0.01$  & 0.51$_{-0.14}^{+0.13}$ & 0.52$_{-0.06}^{+0.03}$  & 5.2$_{-1.2}^{+1.6}$ & 1.22 (65) \\ 
C & 833.7 & 1.0$^\ddagger$ &  0.21--0.89 & 0.32$_{-0.02}^{+0.03}$  & 0.30$^\ddagger$ & 0.50$^\ddagger$  & 1.9$\pm0.3$ & 1.01 (50) \\ 
D & 1537.2 & 1.0($>$0.8) &  0.57--2.05 & 0.30$\pm0.01$  & 0.22$_{-0.04}^{+0.06}$ & 0.44$_{-0.05}^{+0.04}$  & 10.7$_{-2.5}^{+2.1}$ & 1.83 (74) \\ 
E & 523.7 & 1.0$^\ddagger$ &  0.41--1.64 & 0.28$_{-0.03}^{+0.02}$  & 0.30$^\ddagger $ & 0.50$^\ddagger$  & 3.3$_{-0.5}^{+0.8}$ & 0.67 (48) \\ 
F & 871.7 & 1.0$^\ddagger$ &  0.14--0.57 & 0.32$\pm0.06$  & 0.30$^\ddagger$ & 0.50$^\ddagger\ddagger$  & 1.0$_{-0.2}^{+0.4}$ & 0.60 (40) \\ 
G & 1527.0 & 1.0($>$0.8) &  0.46--1.48 & 0.31$\pm0.01$  & 0.32$_{-0.07}^{+0.24}$ & 0.49$_{-0.05}^{+0.03}$  & 6.5$_{-2.7}^{+1.4}$ & 1.48 (65) \\ 
H & 898.8 & 1.0$^\ddagger$ &  0.16--0.58 & 0.30$^\ddagger$  & 0.30$^\ddagger$ & 0.50$^\ddagger$  & 1.2$\pm0.1$ & 1.15 (36) \\ \hline
\end{tabular}
  \label{tab:spec_phase} 
  \end{center}
\footnotemark[${*}$]{In the unit of deg$^{2}$}, 
\footnotemark[${\dagger}$]{X-ray absorption strength relative to that 
assuming Galactic absorption (see text)}, 
\footnotemark[$\ddagger$]{Distributions of the Galactic HI column density 
in terms of 5th to 95th percentiles, in units of 10$^{21}$ atoms cm$^{-2}$},
\footnotemark[$\S$]{Plasma temperature in the unit of keV}, 
\footnotemark[$\parallel$]{Metallicity relative to 
the Ssolar value (Z$_{\odot}$)}, 
\footnotemark[$\#$]{Metallicity ratio of Iron relative to that of others}, 
\footnotemark[${**}$]{Emission Measure in the unit of 10$^{-2}$ cm$^{-6}$ pc},
\footnotemark[$\ddagger$]{Parameter fixed at this value}. 
\end{table*}

In a similar way, we analyzed the spectra from the other Areas 
(figure\,\ref{spec_fig}).
As summarized in table\,\ref{tab:spec_phase}, 
the fits were mostly acceptable,
and the obtained parameters are relatively similar among these Areas.
We examine these results in subsection \ref{subsec:hotplasma}.

\section{Discussions}
\label{sec:discussions}

By accumulating the MAXI/SSC data 
over a large number ($\sim 10^4$)
of ISS revolutions for 2 years,
we have derived soft X-ray maps of the sky,
which are free of any gaps,
and has a solid-state energy resolution,
typically $\sim$ 150 eV (FWHM) at 5.9 keV \citep{2010PASJ...62.1371T}.
We defined an off-source sky region,
and estimated the NXB using the data taken 
when a given  detector area was observing that sky region.
By subtracting the NXB estimated in this way, 
as well as the modeled CXB and the GH,
the final soft X-ray sky maps have been derived
as in figure~\ref{SSC_band_map} and figure~\ref{rgbmap}.
In addition to point sources,
figure\,\ref{rgbmap} reveals the Galactic ridge emission in blue,
and the EXS in red.
The \Lend-1.0\,keV SSC map shows a good similarity,
including the EXS, to the ROSAT map in a similar energy band.

\subsection{Possible contribution by the SWCX component}
\label{subsec:discuss_swcx}

One of the objectives of the present study has been 
to observe the ROSAT-discovered EXS  under a condition 
wherein the variable and anisotropic SWCX contribution 
(\S\ref{bgcomponents}) is considered to be 
much lower than in the ROSAT R5 band (0.56-1.21 keV),
where the SWCX contribution is estimated \citep{2016ApJ...829...83U}
to be $6\% \pm 4\%$ (statistical) $\pm 4\%$ (systematic).
The present SSC results are expected to satisfy this requirement,
for the following reasons.
\begin{enumerate}
\item  We accumulatd the SSC data over 2009-2011,
when the Solar activity was much lower 
(\S\ref{bgcomponents}) 
than in the 190--1991 Solar maximum 
when the RASS was conducted.
\item The SSC data was utilized in $>\Lend $ keV
(\S\ref{bgcomponents}),
where few SWCX lines are present,
and the contribution from the stronger low-energy lines 
should be much smaller thanks to the better 
energy resolution of the SSC than that of ROSAT.
\item 
The SWCX emission is considered strongest in the cusp regions 
above the geomagnetic poles (\S\ref{subsec:discuss_swcx}).
However, SSC-Z never sees the cusps, 
and SSC-H for less than 10\% of its observation time
because the SSC acquires the data only during night times.
\item
During the two years, each sky region 
was in the SSC FOV typically for 150 days,
and was meanwhile scanned by the SSC more than 10$^3$ times.
Therefore, any sporadic count increases due to the SWCX
must have been strongly averaged out.
\end{enumerate}

To examine whether the SWCX contributions is really very low in the SSC data,
we re-calculated the \Lend--1.0 keV map
by removing those time periods
when the Solar proton flux measured with GOES (provided by NOAA)
\footnote{https://www.swpc.noaa.gov/products/goes-proton-flux} 
was$\ge 4\times 10^8$ cm$^{-2}$ s$^{-1}$.
Then, the live time was reduced by 6\%.
When this map is compared with figure~\ref{SSC_band_map}(c)
at each sky direction,
we do not find any difference 
that is higher than $3\times10^{-4}$ counts deg$^{-2}$ cm$^{-2}$ s$^{-1}$
in the surface brightness.
This gives a {\it post facto} 
support to the above expectation
that the SSC results are little contaminated by the SWCX signals.

Another {\it post facto} confirmation is available from figure~\ref{SSC_ROSAT}.
If the SWCX contribution was significant,
we would have observed some spatial differences between MAXI and ROSAT,
because it would be very implausible 
that they are equally contaminated by the SWCX signals.
In particular, we should have observed those data points
where the ROSAT measurements are higher than the SSC results.
However, except for bright variable point sources, 
the SSC and ROSAT measurements are very well correlated 
(figure~\ref{SSC_ROSAT}, \S\ref{subsec:comparison}),
at least for data points 
which are higher than $> 3\times10^{-4}$ counts deg$^{-2}$ cm$^{-2}$ s$^{-1}$.
%$> 3 \times 10^8$ cm$^{-2}$ s$^{-1}$.
(The opposite, i.e., the SSC brightness exceeding that with ROSAT,
is due to the poor spatial resolution of the SSC as mentioned in \S\ref{subsec:comparison}.)
The slight offset, of which the origin is at present unknown,
is in the opposite sense because the ROSAT values are lower.
Therefore, we conclude that not only the SSC map,
but also the ROSAT map, are relatively free from the SWCX signals,
and hence the EXS detected in both these maps
must be of celestial origin.

\subsection{Morphology of the EXS}

As already revealed with ROSAT \citep{1995ApJ...454..643S},
the brightest part of the EXS,
Area B in figure\,\ref{regmap},
forms an arc-like structure,
emerging from the Galactic plane at $\ell\sim 30^\circ$,
and extending along the North Polar Spur.
The EXS brighter than 
$7\times10^{-4}$ counts deg$^{-2}$ cm$^{-2}$ s$^{-1}$
continues via Area A, forming a large circular structure
which almost reaches the north pole ($b\geq 60^\circ$),
and comes back (Areas C and D) 
to the Galactic plane at $\ell\ \sim 300^{\circ}$.
Furthermore, this semi-circular structure
appears to connect to another curved feature
just south of the Galactic plane, 
which runs through Area G and returns 
to the Galactic plane at $\ell\sim 20^\circ$.
The overall structure (Areas B, A, C, D, and G)
thus draws a circle of radius $\sim 50^\circ$, 
centered at  $(\ell, b) \sim (340^\circ, 15^\circ$).

In addition to the main structure of the EXS as described above,
we notice in figure~\ref{SSC_band_map}(e) and figure~\ref{rgbmap}
another, fainter, and thin feature, mainly in the southern hemisphere.
It follows the main feature through Area B, A, C, and D,
but bifurcates 
from it at $\ell\ \sim 300^{\circ}$,
continues down into the south hemisphere 
through the western edge of Area D into H, 
and  reaches $b \sim -70^\circ$.
It then comes back, through Areas F and E,
to the Galactic plane at $\ell\ \sim 30^{\circ}$.
This structure thus draws a hollow ellipse,
which is approximately centered on the GC 
and elongated from north-north-west to south-south-east.
Although this feature was  probably not noticed before,
it is visible also in the ROSAT map (figure~\ref{SSC_band_map}f),
which reinforces its detection with the SSC.

Thus, except the much higher brightness along the North Polar Spur,
the overall morphology of the EXS 
has an approximate north-to-south symmetry.
Compared to these EXS distributions,
the Fermi Bubble is considerably smaller in size,
and appears to be sitting inside the EXS structure.
Only in Area B, the outer edge of the Fermi Bubble 
may be touching the inner edge of the EXS,
as pointed out by \citet{2013ApJ...779...57K}.

%% Fermi Bubbleとの関係は、こんなもので良いですか？

\subsection{Thin hot plasma of the EXS}\label{subsec:hotplasma}

The EXS clearly shows thermal spectra (\S \ref{subsec:specana}),
as already indicated by the ROSAT observations \citep{1995ApJ...454..643S}.
As summarized in table\ref{tab:spec_phase}, 
the temperatures in the eight Areas 
(except H where the brightness is too low)
are all consistent, within errors, with 0.31 keV.
In addition, in B, D, and G where the brightness is rather high,
the fit yielded $Z = 0.2-0.5~\zo$ and $\zfe/Z_{o} \sim 0.5$.
Although we were unable to constrain $Z$
in Areas A, C, E, F and H,
due to the low surface brightness
the spectra of these fainter Areas are consistent
also with $Z = 0.3~\zo$ and $\zfe/Z_{o} = 0.5$.
These spectral results approximately agree 
with those obtained by Suzaku pointing observations 
onto very limited and narrow regions of the EXS
\citep{2013ApJ...779...57K, 2015ApJ...802...91T, 2018ApJ...862...88A}.

We tried to measure the distance to the bright EXS Areas,
by analyzing the absorption feature at low energy. 
The low energy emission was found to be absorbed by $\fgal>$0.6. 
A more detailed analysis at lower galactic latitude  below 10 degrees 
or at low energy band may give us a hint to solve this problem 
and left as a future work.

Energy spectral shapes of fainter areas were expressed 
by almost the same spectral parameters with those of bright areas.
The entire structure (bright and faint areas) was added 
up to form a large elliptical shape centered at the GC.
More detailed analysis on faint areas such as the significance 
of southern structure or distribution of GH in the sky, 
will be reported elsewhere.

Since X-ray spectrum shows a thin thermal origin, 
we can measure the average EM of the large structure along the line of sight. 
If we assume that the large structure is a spherical shape, 
we can estimate the mass, $M_\mathrm{t}$ and 
the thermal energy $E_\mathrm{th}$ contained 
where $D$ is the distance to the structure.  
In this calculation, we assume a uniform density for simplicity.
\begin{eqnarray*}
M_\mathrm{t} &\sim & 5.8\times 10^7 \left(\frac{EM}{15\times10^{-3}\,\mathrm{cm}^{-6}\,\mathrm{pc}} \right)^{0.5} \, \left(\frac{D}{8\,\mathrm{kpc}}\right)^{2.5}~ M_{\odot}\\
E_\mathrm{th} &\sim & 3\times 10^{55} 
\left(\frac{kT}{0.31~\mathrm{keV}} \right) 
\left( \frac{D}{8~\mathrm{kpc}} \right)^{2.5} ~\mathrm{erg}
\end{eqnarray*}

\subsection{Possible origin of the EXS}\label{subsec:exsorigin}

If the X-ray large structure is centered on the galactic center, we can estimate $\mathrm{D}\sim 8\,\mathrm{kpc}$.  Furthermore, apparent temperature looks uniform over the large structure.  This requires us a very strong AGN activity in our galaxy as well as the heat mechanism to keep the temperature constant at the far distant from the GC.  Even if it is out of the galactic plane, it can be a local structure and close to us.  The value of $\mathrm{D}$ can be 1\,kpc.  Then, the mass and the total energy will be reduced a lot that could be generated by a hypernova \citep{1998Natur.395..672I} or a series of supernova \citep{2013PASJ...65...14K}.  The detailed discussion will be done after obtaining a finer spatial resolution and better statistics in near future.

\section{Conclusion}

We analyzed the 0.7--7 keV soft X-ray all-sky data,
acquired with the MAXI/SSC
during the first two years of its continued operation on the ISS.
Carefully removing backgrounds, including the NXB in particular,
we obtained NXB-subtracted all-sky maps in three energy bands.
The high (2--4\,keV) and medium (1--2\,keV) energy maps 
show many point sources as well as the galactic ridge emission,
in agreement with those from the GSC on MAXI and other satellites.  
The low energy map, in \Lend--1\,keV, reveals the EXS
in addition to many point sources.  
The EXS brightness distribution measured with the SSC is fully consistent 
with that observed by ROSAT in a corresponding energy range,
in spite of the 20 years of interval.
Further taking into account the lower Solar activity,
the higher energy resolution, 
and observation geometry of the SSC,
we conclude that the EXS observed with the SSC 
in the  \Lend--1\,keV range is a celestial phenomenon,
with little contamination by the SWCX.
This also applies to the ROSAT R5 band map.
The entire outline of the EXS looks like an ellipse centered at the GC,
and its brighter part  is well within a circle of about 50$^\circ$ radius 
centered around ($\ell\sim 340^\circ, b\sim 15^\circ$).  
The EXS spectra are relatively similar 
among these regions of different brightness,
and can be interpreted as thin thermal emission
with a temperature of 0.31\,keV  and depleted abundances.
The interstellar absorption affecting the EXS was constrained as 
$> 0.6$ times the Galactic absorption column in the corresponding direction.
Therefore, the distance to the hot plasmas 
that emit the EXS still remains uncertain.
If the EXS emitter is located at a distance similar to that of the GC,
it may be related to some past activity of our GC.
If, instead, it is considerably closer to us than the GC,
a hypernova or a series of supernovae may be considered
as a plausible origin.
Future X-ray observations below 0.7\,keV will give 
a further clue to its origin.

\begin{ack}
This research has made use of MAXI data provided by RIKEN, JAXA and the MAXI team. 
\end{ack}

\bibliographystyle{apj}
\bibliography{sscdiffuse}

\end{document}